\documentclass[11pt]{article}

\usepackage[top=1in, left=1in, right=1in,
bottom=1in]{geometry}

\usepackage{latexsym}
\usepackage[parfill]{parskip}
\usepackage{amssymb,amsmath, bm}
\usepackage{graphicx}
\usepackage{marvosym}
\usepackage{lscape}
\usepackage{rotating}
\usepackage{setspace}
\usepackage{threeparttable}
\usepackage{booktabs}
\usepackage{multirow}

\usepackage{caption}
\usepackage{subcaption}
\usepackage{comment}
\usepackage{bbm}
\usepackage{subcaption}

\usepackage{algorithm}
\usepackage{algpseudocode}

\usepackage{tikz}
\usetikzlibrary{positioning,matrix}

\usepackage{natbib}

\usepackage{color}
\usepackage[pdftex, bookmarksopen=true, bookmarksnumbered=true,
pdfstartview=FitH, breaklinks=true, colorlinks = true, citecolor = blue, linkcolor = blue]{hyperref}
 \usepackage{hyperref}

\usepackage{dcolumn}
\newcolumntype{.}{D{.}{.}{-1}}
\newcolumntype{d}[1]{D{.}{.}{#1}}

\usepackage{amsthm}

\newcommand\independent{\protect\mathpalette{\protect\independenT}{\perp}}
\def\independenT#1#2{\mathrel{\rlap{$#1#2$}\mkern2mu{#1#2}}}

\usepackage{color-edits}

\newcommand{\E}{\ensuremath{\mathbb{E}}}

\newcommand{\bK}{\mathbf{K}}

\newcommand{\rf}{\text{rf}}
\newcommand{\bart}{\text{bart}}

\def\b1{\boldsymbol{1}}

\usepackage{xcolor}

\begin{document}

\pagestyle{plain}

\newcommand{\blind}{0}

\newcommand{\tit}{Forest Kernel Balancing Weights:\\Outcome-Guided Features for Causal Inference\thanks{This research is supported by the Institute of Education Sciences, U.S. Department of Education, through Grants R305D210014 and R305D240036. Andy Shen is supported by the National Science Foundation (NSF) Graduate Research Fellowship under Grant No. 2146752. The opinions expressed are those of the authors and do not represent views of the funding institutions. }}

\if0\blind

{\title{\vspace{-1em}\tit}
\author{Andy A. Shen\thanks{University of California, Berkeley, Email: aashen@berkeley.edu}
\and Eli Ben-Michael\thanks{Carnegie Mellon University, Email: ebenmichael@cmu.edu}
\and Avi Feller\thanks{University of California, Berkeley, Email: afeller@berkeley.edu}
\and Luke Keele\thanks{University of Pennsylvania, Email: luke.keele@gmail.com}
\and Jared Murray\thanks{University of Texas, Austin, Email: jared.murray@mccombs.utexas.edu}
}

\date{\today}

\maketitle
}\fi

\if1\blind
\title{\bf \tit}
\maketitle
\fi

\newenvironment{revision}{\color[HTML]{000000}}{\ignorespacesafterend}
\newcommand{\smallrevision}[1]{\textcolor[HTML]{000000}{#1}}

\begin{abstract}
While balancing covariates between groups is central for observational causal inference, selecting which features to balance remains a challenging
problem. Kernel balancing is a promising approach that first estimates a kernel that captures similarity across units and then balances a (possibly low-dimensional) summary of that kernel, indirectly learning important features to balance.
In this paper, we propose \textit{forest kernel balancing}, which leverages the underappreciated fact that tree-based machine learning models, namely random forests and Bayesian additive regression trees (BART), implicitly estimate a kernel based on the co-occurrence of observations in the same terminal leaf node.
Thus, even though the resulting kernel is solely a function of baseline features, the selected nonlinearities and other interactions are important for predicting the outcome---and therefore are important for addressing confounding.
Through simulations and applied illustrations, we show that forest kernel balancing leads to meaningful computational and statistical improvement relative to standard kernel methods, which do not incorporate outcome information when learning features.
\end{abstract}


\clearpage
\onehalfspacing

\section{Introduction}
A central goal of observational study design is to achieve \textit{covariate balance}, which ensures that treated and control groups are comparable on observed characteristics \citep{benmichael2021_review}. 
Weighting methods accomplish this by, for example, re-weighting the control group to have a comparable covariate distribution to the treated group, such as via inverse propensity score weighting (IPW). 
However, choosing which features to re-weight can be challenging: failing to capture important nonlinearities or other complex relationships can lead to bias.

\emph{Kernel balancing} is a promising data-driven approach that uses kernel methods to indirectly learn a (possibly infinite) set of features to balance \citep{wong2018kernel,hazlett2020kernel,kim2024scalable}. The key idea is to first compute a kernel matrix $\mathbf{K}$ and then use existing weighting approaches to balance a low-dimensional summary of $\mathbf{K}$.
Kernel methods are typically \emph{design based}: they encode similarity between units based on a fixed kernel derived from baseline features only. While such kernels can capture complex relationships between covariates, they may fail to reflect nonlinearities and other relationships that are important for predicting the outcome---and therefore that are important for addressing confounding. 

In this paper, we propose \emph{forest kernel balancing}, in which we use outcome information to learn the kernel matrix $\mathbf{K}$ to balance. In particular, we leverage the fact that tree-based machine learning models, namely random forests (RF) and Bayesian additive regression trees (BART), implicitly estimate a kernel based on the co-occurrence of observations in the same terminal leaf node \citep{breiman2000some,scornet2016random,hill2020bayesian}. Thus, while $\mathbf{K}$ is still solely a function of the baseline features, incorporating the outcome into learning $\mathbf{K}$ ensures that these relationships are pertinent to the outcome, which appropriately captures more sources of confounding \citep{jin2025cross}.
We conduct extensive simulation studies and show that forest kernel balancing reduces bias and RMSE compared to design-based kernel existing methods. 
We then demonstrate this approach in practice with an applied illustration to an educational training program \citep{keller2025new} and an observational study of the effects of child soldiering \citep{blattman2010consequences}. 

This paper builds on an extensive literature on kernel methods implied by random forests and related tree-based methods, as well as an emerging literature on leveraging outcome information to improve observational study design \citep{jin2025cross,de2025data}. See \citet{breiman2000some} for an early connection and \citet{scornet2016random,balog2016mondrian} for more detailed discussion. Separately, \citet{lin2006random} provide an early interpretation of random forests as a weighted nearest neighbor methods. For BART, the connection to kernel methods is typically via BART's representation as a Gaussian process; see \citet{hill2020bayesian} for a review.

Our paper is structured as follows: Section \ref{sec:background} provides our setup and background on different weighting approaches. Section \ref{sec:methodology} introduces forest kernels and kernel balancing.
Section \ref{sec:simulation-study} compares performance via a series of simulations.
Section \ref{sec:application} demonstrates the performance of our kernel with two empirical case studies. Section \ref{sec:conclusion} concludes. The code to reproduce our findings can be found at the following link: \href{https://github.com/aashen12/forest-kbal}{https://github.com/aashen12/forest-kbal}.


\section{Background} \label{sec:background}

\subsection{Notation, assumptions, and estimands}
We assume we have a sample of units $i=1,\dots,n$, drawn i.i.d. from some population. We denote a binary treatment $Z_i$, where $Z_{i} = 1$ and $Z_i = 0$ corresponds to the treated and control conditions, respectively; $n_1$ and $n_0$ denote the corresponding number of treated and control units. We let $Y_i$ denote the observed outcome, and we use the potential outcomes framework, where each unit has two potential responses $Y_{i}(1)$ and $ Y_{i}(0)$, corresponding to their respective response under treatment and control. The observed outcome is $Y_i = Z_iY_i(1) + (1 - Z_i) Y_i(0)$. Finally, we invoke the Stable Unit Treatment Value Assumption \citep{Rubin:1986}, which implies no hidden forms of treatment and no interference across units.
To reduce notational complexity, we will drop the $i$ subscript when taking expectations and probabilities. We use $X_i$ to denote a vector of baseline covariates. Throughout, we will make use of the \emph{propensity score}, the probability of receiving treatment conditional on the baseline covariates, which we denote as: $e(x) = \mathbb{P}(Z = 1 \mid X = x)$. 

Our estimand of interest is the \textit{average treatment effect on the treated (ATT)}, the average difference in outcomes for those individuals in the population actually exposed to treatment:\footnote{While our work focuses on the ATT for clarity, the proposed methodology can be extended to other common estimands, such as the average treatment effect (ATE) or the average treatment effect for the overlap population (ATO).}
\begin{align}
\label{eqn:att}
    \tau = \mathbb{E}\left[Y(1) - Y(0) \mid Z = 1\right] =  \mathbb{E}\left[Y(1) \mid Z = 1\right] -  \mathbb{E}\left[Y(0) \mid Z = 1\right].
\end{align}
For individuals exposed to treatment, we directly observe $Y(1)$ and therefore can directly estimate the mean outcome under treatment for this group, $\mu_1 \equiv \mathbb{E}[Y(1) \mid Z = 1]$, such as via the sample average outcome among the treated. The challenge is then to estimate the average counterfactual outcomes for this group, $\mu_0 \equiv \mathbb{E}[Y(0) \mid Z = 1]$, which will be the focus of our discussion.

We require two additional assumptions to identify $\mu_0$. First, we assume that treatment assignment is ignorable, conditional on the covariates $X$ (i.e., no unmeasured confounding):
\[
Y(0), Y(1) \independent Z \mid X.
\]
Second, we assume there is overlap in the covariate distributions between the treated and control groups: for the ATT, this assumption requires that no units are deterministically treated, i.e. $e(x) < 1$ for each $x$ in the population. 

Under these assumptions, the average counterfactual outcome $\mathbb{E}[Y(0) \mid Z = 1]$ (and therefore the ATT) is identified as a function of the propensity score and observed outcomes:
$$
\mathbb{E}[Y(0) \mid Z = 1] = \mathbb{E}\Bigg[\underbrace{\frac{e(X)}{1 - e(X)}}_{\text{IPW}}Y \mid X, Z = 0\Bigg],
$$
where $e(x)/(1-e(x))$ are the ``inverse'' propensity score weights. 

\subsection{Estimating inverse propensity score weights via balancing} \label{sec:bal-wgt-intro}
Our goal is to construct a weighting estimator for the mean counterfactual outcome for the treated group, $\mu_0$. With known propensity score $e(x)$, we can estimate this unobserved portion of the ATT via a  weighted average of the outcomes in the control group:
\begin{align}
\label{eqn:tauhat}
\hat{\mu}_0 = \sum_{i=1}^n (1 - Z_i)w_iY_i.
\end{align}
Based on the identification formula above, we should set $w_i$ to equal the inverse propensity score weights, $w_i =  \frac{e(x_i)}{1 - e(x_i)}$.
The key challenge is to estimate these inverse propensity score weights when the propensity score is not known.
Traditional IPW first estimates the propensity score $e(x)$ and then plugs the estimated values $\hat{e}(x)$ into the analytic form for the weights. 
This approach, however, can lead to unstable estimates and poor implied covariate balance with even moderate dimensional covariates $x$: since $e(x)$ is in the denominator, errors in $\hat{e}(x) - e(x)$ can cause the estimator to ``blow up'' \citep[see][]{benmichael2021_review}.

Instead we focus on methods that estimate inverse propensity score weights directly. To motivate this approach, note that under standard identifying conditions we can rewrite the observed outcome for unit $i$ as
$$
Y_i = (1-Z_i)Y_i(0) + Z_iY_i(1) = \mu_0(x_i) + Z_i\tau_i + \varepsilon_i,
$$
where $$\mu_0(x) \equiv \E[Y(0)\mid X=x] = \E[Y\mid X=x, Z=0]$$ is the potential outcome function; $\tau_i \equiv Y_i(1) - Y_i(0)$ is the individual-level treatment effect; and $\varepsilon_i$ is conditionally mean zero noise. We can then write a weighting estimator of the unobserved counterfactual mean $\mu_0$:
\begin{equation} \label{eqn:bias-bound}
\hat\mu_0= \sum_{i=1}^n(1-Z_i)w_iY_i = 
\frac{1}{n_1} \sum_{i=1}^nZ_i\mu_0(x_i)+ 
\underbrace{\left(\sum_{i=1}^n(1-Z_i)w_i\mu_0(x_i) - \frac{1}{n_1}\sum_{i=1}^nZ_i\mu_0(x_i)\right)}_{\text{imbalance in }\mu_0(x)} +~\text{noise},
\end{equation}
where we add and subtract $\frac{1}{n_1} \sum_{i=1}^nZ_i\mu_0(x_i)$, a sample analogue of the unobserved counterfactual mean, and where the noise term is a mean-zero weighted sum of the $\varepsilon_i$ terms. 
The core task of a weighting estimator is then to control the imbalance term. If we knew the true potential outcome function $\mu_0(x)$, we could simply find weights to balance this quantity. 

\paragraph{Balancing weights.}
Unfortunately, we do not know $\mu_0(x)$. Instead, we can leverage an important property of inverse propensity score weights: they are the unique weights that balance \textit{any} (integrable) function between the treated and control groups, including the unknown $\mu_0(x)$.
The space of all integrable functions is very large, and so balancing weights estimators posit a \textit{function class} for $\mu_0(\cdot)$ that restricts the model in some way. There is a direct correspondence between the choice of function class and the form of covariates or their transformations that should be balanced \citep{benmichael2021_review}. As a leading example, if $\mu_0(x) = x^{\top}\beta$ is linear in the covariates $x$, then the resulting imbalance term is controlled by the difference between the average value of the covariates $x$ in the treated group and the \emph{weighted} average in the control group.

In practice, it is useful to consider weights that achieve \textit{approximate} balance and also penalize the dispersion of the weights. 
Following the developments in \citet{benmichael2021_lor,benmichael2021_drp,keele2023hospital,keele2025balancing},
we consider the following ``linear'' balancing weights objective function:
\begin{align}
\label{eqn:bal-wgt}
\min_{w \in \Delta^n} \underbrace{\left\lVert  \sum_{i=1}^n (1-Z_i) w_i \mathbf{X}_i - \frac{1}{n_1} \sum_{i=1}^n Z_i \mathbf{X}_i \right\rVert^2_2}_{\text{imbalance}} + \lambda \underbrace{\sum_{i=1}^n w_i^2}_{\text{dispersion}},
\end{align}
where we pre-process each covariate $X_k$ to have mean 0 and variance 1. We can identify this objective as representing a bias-variance tradeoff, where the leading term represents worst-case confounding over functions linear in $x$ and the penalty reduces the dispersion of the weights \citep{benmichael2021_lor,keele2025balancing,Hirshberg2019_amle,ben2023using}. Following \citet{Hirshberg2019_amle} and \citet{ben2023using}, we set this hyperparameter to the residual variance from a regression of the outcome (scaled to have mean zero and variance one) on covariates within the control group.
We also restrict the weights to be on the simplex $w \in \Delta^{n}$ (i.e., they are non-negative and sum to 1), though we can relax this restriction \citep{arbour2025regularizing}.\footnote{Note that in this formulation, weights on treated units are also part of the vector $w$, but because they  do not appear in the imbalance objective they are regularized to zero if $\lambda > 0$.}


Different approaches to balancing weights make different choices about these functions, such as  replacing the 2-norm of imbalance with other norms, though they all follow the same principle of minimizing a combination of imbalance and weight dispersion/complexity; see \citet{zubizarreta2015stable,Wang2020,athey2018approximate,tan2020regularized}. 
%
\begin{revision}
Finally, the covariate balancing propensity score (CBPS) of \citet{imai2014covariate} combines a logistic regression propensity score model and balancing weights by incorporating balance constraints into the estimation of the propensity model itself \citep{fong2018covariate,fan2022optimal}.
Since the innovation in our forest kernel approach is feature construction rather than the mechanics of the weighting procedure itself, we can easily incorporate alternative weighting approaches, such as via estimating CBPS using our learned features. We evaluate this directly in Section \ref{sec:cbps} of the supplement, finding that CBPS with learned features improves performance relative to CBPS with the original features alone.
\end{revision}

\paragraph{Feature maps.} 
Other model classes induce additional balance terms, e.g. positing that $\mu_0(\cdot)$ is a quadratic model corresponds to measuring imbalance across all covariates and their second-order interactions.
Generally, we can denote this via a {\em feature vector} $\phi: \mathcal{X}\rightarrow \mathbb{R}^q$:
$$
X\rightarrow \phi(X). 
$$
For example, if $\phi$ includes quadratic terms, then the balancing objective aims to balance the first two moments of the covariates.

The goal of choosing features $\phi(x)$ is to construct a rich set of functions $\phi(x) = (\phi_1(x), \phi_2(x), \dots)$ that is sufficiently large so that it can adequately represent $\mu_0(\cdot)$, i.e. $\mu_0(x)\approx \phi(x)^\top\beta$, but not so large that achieving good balance in a finite sample is too difficult \citep{benmichael2021_review}; see also \citet{austin2007comparison,austin2015moving} for extensive discussion.\footnote{An alternative motivation for selecting $\phi(x)$ is to instead find functions that adequately represent the true inverse propensity score weights themselves. See, for example, \citet{jin2025cross}.}
Given this feature map $\phi$ we can then consider a modified balancing weights objective: 
\begin{align}
\label{eqn:bal-wgt-features}
\min_{w \in \Delta^n} \underbrace{\left\lVert  \sum_{i=1}^n (1-Z_i) w_i \phi(\mathbf{X}_i) - \frac{1}{n_1} \sum_{i=1}^n Z_i \phi(\mathbf{X}_i) \right\rVert^2_2}_{\text{imbalance in }\phi(x)} + \lambda \underbrace{\sum_{i=1}^n w_i^2}_{\text{dispersion}}.
\end{align}

Once again we have an issue: selecting appropriate features $\phi(x)$ is challenging because we do not know $\mu_0(x)$. Here we can construct the feature map by hand, or, as we discuss next, implicitly by specifying a kernel function.

\subsection{Kernels and kernel balancing} \label{sec:kbal}

\textit{Kernel balancing} is a generalization of balancing weights that minimizes an objective like~\eqref{eqn:bal-wgt-features} with a (potentially) infinite number of features, using the ``kernel trick'' to avoid explicitly forming the features. We briefly review kernels and kernel balancing here, focusing on Mercer and generic inner-product kernels, which will cover all of our use cases and most of the kernel balancing literature. 

A {\em kernel} is a positive definite function $k(\cdot, \cdot): \mathcal{X} \times \mathcal{X} \mapsto \mathbb{R}$ that encodes similarity between observations based on their covariates. By the reproducing property:
$$
k(x_i, x_j) = \langle k(\cdot, x_i), k(\cdot, x_j) \rangle_{\mathcal{H}},
$$
where the inner product is in the reproducing kernel Hilbert space (RKHS) associated with the kernel $\mathcal{H}$. 
For many common kernels, we can make these features explicit by choosing an infinite-dimensional map $\phi(x): \mathcal{X}\rightarrow \mathbb{R}^\infty$ such that
$$
k(x_i x_j) = \left\langle \phi(x_i), \phi(x_j)\right\rangle_2 = \sum_{h=1}^\infty \phi_h(x_i)\phi_h(x_j).
$$
We can therefore think of a kernel as encoding a (non-unique) transformation of the original covariates, analogous to augmenting our original variables with polynomial expansions, interactions, or other transformations.

The famous \textit{kernel trick} then allows us to substitute $\phi(x)$ for $x$ in the imbalance component of the objective function in Equation~\eqref{eqn:bal-wgt}, even with infinite-dimensional $\phi(x)$, without ever explicitly forming the feature vectors $\phi(x)$ themselves. 
For any two $a_0, a_1\in\mathbb{R}^n$ we have
\begin{align*}
\left\langle \sum_i a_{0i} \phi(x_i), \sum_j a_{1j} \phi(x_j)\right\rangle_2 &= \sum_{i,j} a_{0i}a_{0j} \phi(x_i)^\top\phi(x_j) = \sum_{i,j} a_{0i}a_{0j} k(x_i, x_j).
\end{align*}
We can then write the imbalance term in Equation \eqref{eqn:bal-wgt-features} as
\begin{gather*}
{\left\lVert  \sum_{i=1}^n (1-Z_i) w_i \phi(\mathbf{X}_i) - \frac{1}{n} \sum_{i=1}^n Z_i \phi(\mathbf{X}_i) \right\rVert^2_2} = a_0^{\top}\mathbf{K}a_0 + a_1^{\top}\mathbf{K}a_1 - 2a_0^{\top}\mathbf{K}a_1,\\
a_{0i} = (1-Z_i)w_i,\quad a_{1i} = Z_i/n.
\end{gather*}
This is 
 (up to scaling and regularization) 
an instantiation of {\em kernel mean matching} \citep{gretton2009covariate}, where we learn weights that shift the distribution of features in the controls to match the distribution of features among treated units.

\paragraph{Approximate kernels and spectral truncation.} 
While many kernels implicitly induce infinite-dimensional feature vectors, in practice at most $n$ of these are realized in any given sample (and typically the effective dimension is much lower). That is,
$$
\mathbf{K} = \Phi\Phi^\top \approx \hat\Phi\hat\Phi^\top,
$$
where $\hat\phi(x)$ is an approximate feature vector of length $r \ll q$.
An important special case of this is the \textit{spectral approximation}; see \citet{wong2018kernel,hazlett2020kernel,kim2024scalable,hartman2024kpop}. Let $\mathbf{K} = \mathbf{U}\mathbf{D}\mathbf{U}^\top$ be the eigendecomposition of the kernel matrix, where $\mathbf{D} = \text{diag}(\sigma^2_1, \ldots, \sigma^2_n)$. The spectral approach retains the top $r$ principal components of $\mathbf{K}$, scaled by their singular values: $\hat\phi_k(x) = \sigma_k U_{k}$ for $k = 1, \ldots, r$. We can then minimize imbalance in these (scaled) principal components directly. Imbalance in the trailing principal components (i.e., $k = r+1, \ldots, n$) is small in absolute value because the corresponding singular values are small. As a result, using the low-rank approximation $\hat\Phi$ instead of $\Phi$ can significantly enhance computational efficiency, and potentially enhance statistical efficiency, with only minimal additional imbalance.

Choosing $r$ is therefore an important hyperparameter with many kernel balancing methods. 
To do so, \cite{hazlett2020kernel} defines a ``worst-case'' bias bound and chooses $r$ to minimize this bias bound---but also demonstrates that estimation accuracy is relatively insensitive to this choice, so long as $r$ is not too extreme. 
We report results over a grid of $r$ values and proceed conditionally on the selected value, noting that any conventional selection rule can be used and the subsequent weighting and estimation steps are unchanged.

\paragraph{Kernel choices.} The last key ingredient for kernel balancing is specifying the kernel itself, typically in addition to the spectral truncation considered above. We focus on two common choices: \textit{polynomial} and \textit{Gaussian} kernels.

In a polynomial kernel, the features are simply a polynomial expansion of the original covariates:
$$
k(x_i, x_j) = (x_i^\top\Sigma x_j + c)^d,
$$
where $c$ is a constant that balances higher and lower order polynomial terms, $d$ is the degree of the polynomial, and $\Sigma$ is a $p\times p$ matrix (used, for example, to whiten the original variables). Expanding the kernel and collecting terms one can show that the corresponding feature map comprises a $d-$degree polynomial transformation of the original covariates. While this setup is quite general, it is useful to focus on the simplified case of the \textit{linear kernel} with $c=0, d=1, \Sigma = \mathbb{I}$: $k(x_i, x_j) = x_i^\top x_j$. This exactly recovers the linear balancing weights problem introduced above (where we can modify $\Sigma$ to recover standardization).

The most common alternative in kernel balancing is the {\em Gaussian} or {\em radial basis function} kernel \citep{wong2018kernel,hazlett2020kernel,kim2024scalable,hartman2024kpop}:
\begin{align*}
    k(x_i, x_j) = \exp({-\left\lVert x_i - x_j \right\rVert_2^2 / b}),
\end{align*}
where $b$ is the kernel bandwidth (a hyperparameter to be tuned) and $x_i$ and $x_j$ are scaled to have zero mean and variance 1. The Gaussian kernel is more flexible than the linear kernel and can 
approximate a wide class of functions while remaining an intuitive measure of similarity: similar covariate profiles receive kernel values closer to one, and increasing $b$ captures more global patterns of similarity. In terms of the implied feature map $\phi(x)$, increasing $b$ puts higher weight on smooth, low-frequency features while decreasing $b$ puts higher weight on more irregular functions. 
There are many approaches for choosing bandwidth $b$, such as the value that produces maximal variance in $\bK$ \citep{hartman2024kpop} or the rank of $\mathbf{X}$ \citep{hazlett2020kernel,kim2024scalable}.

\paragraph{Combining kernels.}
Finally, it is often useful to balance both the raw covariates in addition to their feature space representation; see the  ``mean-first'' procedure from \citet{hartman2024kpop}.
This corresponds to constructing a new kernel:
$$
k(x_i, x_j) = x_i^\top \Sigma x_j + k_0(x_i, x_j),
$$
the sum of a degree 1 (linear) polynomial kernel and another (e.g., Gaussian) kernel $k_0$, either in its original form or using a spectral approximation. 
Practically this combined kernel amounts to modifying the imbalance term in Equation \eqref{eqn:bal-wgt-features} to concatenate the vector of raw covariates (columns of $\mathbf{X}$) and the re-scaled principal components of $\mathbf{K}$, $\sigma_kU_k$.

A key practical consideration is the relative scale of the two sets of features, which implicitly defines their relative importance in the squared distance balance metric used in Equation \eqref{eqn:bal-wgt-features}. As a default, we propose to normalize each ``block'' to have equal overall variance. In particular, we standardize $X$ to have mean 0 and variance $1/p$, where $p$ is the number of raw covariates; then $\sum_k \text{var}(X_k) = 1$. Using the spectral approximation for $\mathbf{K}$, we similarly re-scale the variance of each principal component so that the variances across the $r$ components also sum to one: the variance of principal component $j$ is $\sigma^2_j / \sqrt{\sum_{i=1}^r \sigma^2_i}$ (instead of $\sigma^2_j$). This ``block normalization''  treats the raw covariates and kernel principal components as equally important \textit{in aggregate} while preserving the relative importance of the principal components rather than flattening them. This way, neither the raw covariates nor the kernel features dominate the balancing weight objective function \eqref{eqn:bal-wgt-features} and each kernel feature still maintains its importance in terms of explained variance. In practice, the total variance allocated to each block (raw or kernel) can be determined via an additional hyperparameter. While we choose equal weighting below, any choice is compatible with our framework.

\paragraph{Learning what to balance and the limits of design-based kernels.}

As we discuss in Section \ref{sec:bal-wgt-intro}, the goal of choosing features $\phi(x)$ is to construct a sufficiently rich set of functions $\phi(x) = (\phi_1(x), \phi_2(x), \dots)$ such that we can approximate the true potential outcome function, $\mu_0(x)\approx \phi(x)^\top\beta$.
Choosing $\phi(x)$ via a kernel, however, does not ``solve'' the problem of what to balance: we have simply exchanged hand-selected features for the specification of a kernel family and their parameters, which is typically more opaque. Without strong subject matter knowledge, assessing the trade-offs between kernel choices can be difficult.
Moreover, commonly used kernels, including linear, polynomial and Gaussian kernels can be sensitive to transformations of the data: kernel weights based on similarity in income can be wildly different from weights based on similarity in the logarithm of income. 
The bandwidth in a Gaussian kernel is also not invariant to nonlinear transformations of the inputs. 

More broadly, the kernel balancing framework thus far is fully \textit{design-based}, meaning that the kernel matrix and corresponding weights are constructed using only baseline covariates. Design-based methods are appealing because they mirror the logic of randomized trials in which treatment assignment precedes outcomes. However, features constructed solely from covariate-driven kernels may be insufficient for approximating $\mu_0(x)$. 

Our main proposal, which we turn to next, is to instead use outcome information to learn important features to balance; see also \citet{jin2025cross}. In particular, we propose learning forest-based kernels, coupled with data splitting to preserve valid inference, tuned using outcome data to help learn a relevant feature representation.


\section{Forest Kernel Balancing} \label{sec:methodology}
Our main contribution is to extend kernel balancing by replacing design-based kernels with outcome-guided kernels derived from tree-based machine learning methods, known as \textit{forest kernels}. We consider two popular choices: random forests (RF) and Bayesian additive regression trees (BART). Using outcome data helps target a relevant basis by focusing attention on feature representations that are highly correlated with outcomes. Using tree-based kernels also makes kernel balancing more robust: unlike linear and Gaussian kernels, forest kernels are largely invariant to marginal transformations of the covariates and robust to outlying observations \citep{breiman2001random,lin2006random}. We begin by describing forest kernels in general and the implied kernels of RF and BART in particular. We then turn to broader implementation issues.

\subsection{Forest Kernels} \label{sec:forest-kernels}

Tree ensemble methods like random forests, gradient boosted trees, and Bayesian additive regression trees (BART) implicitly define a similarity metric --- a kernel --- over their input spaces. There is a large literature in machine learning and statistics on such kernels and their properties \citep{breiman2000some,lin2006random,scornet2016random,balog2016mondrian,hill2020bayesian}. We briefly review key results here.

\begin{figure}[tbp]
\centering

\newcommand{\panelscale}{0.7}

\begin{subfigure}{0.32\textwidth}
\centering
\scalebox{\panelscale}{%
\begin{tikzpicture}[
  every node/.style={font=\small},
  treenode/.style={rectangle,draw,rounded corners,
                   minimum width=16mm,minimum height=8mm,
                   inner sep=2pt},
  leaf/.style={circle,draw,minimum size=8mm,inner sep=1pt},
  leafhighlight/.style={leaf,draw=orange!90!black,very thick},
  hledge/.style={orange!90!black,very thick}
]

\node[treenode]        (root) at (0,0)   {$x_1 < 0.2$};
\node[leaf]            (L11)  at (-2,-2) {$L_{1}$};
\node[treenode]        (mid)  at (2,-2)  {$x_3 < 0.7$};
\node[leafhighlight]   (L12)  at (1,-4)  {$L_{2}$};
\node[leaf]            (L13)  at (3,-4)  {$L_{3}$};

\draw (root) -- (L11) node[midway,left] {T};
\draw[hledge] (root) -- (mid) node[midway,right] {F};
\draw[hledge] (mid) -- (L12) node[midway,left] {T};
\draw (mid) -- (L13) node[midway,right] {F};

\end{tikzpicture}%
}
\caption{\label{fig:tree-decision-tree} Decision tree}
\end{subfigure}
\hfill
\begin{subfigure}{0.32\textwidth}
\centering
\scalebox{\panelscale}{%
\begin{tikzpicture}[scale=4,>=stealth, every node/.style={font=\small}]
  \draw[->] (0,0) -- (1.08,0) node[right] {$x_1$};
  \draw[->] (0,0) -- (0,1.08) node[above] {$x_3$};

  \draw (0,0) rectangle (1,1);

  \draw (0.2,0) -- (0.2,1);
  \draw (0.2,0.7) -- (1,0.7);

  \draw (0.2,0) -- (0.2,-0.02) node[below] {$0.2$};
  \draw (0,0.7) -- (-0.02,0.7) node[left] {$0.7$};

  \filldraw[fill=orange!20,draw=orange!90!black,thick]
    (0.2,0) rectangle (1,0.7);

  \node at (0.1,0.35) {$L_{1}$};
  \node[text=orange!90!black] at (0.6,0.35) {$L_{2}$};
  \node at (0.6,0.85) {$L_{3}$};
\end{tikzpicture}%
}
\caption{\label{fig:tree-partition} Induced partition}
\end{subfigure}
\hfill
\begin{subfigure}{0.32\textwidth}
\centering
\scalebox{\panelscale}{%
\begin{tikzpicture}[>=stealth, every node/.style={font=\small}]

\matrix (X) [matrix of math nodes,
             ampersand replacement=\&,
             left delimiter={[}, right delimiter={]},
             row sep=3pt, column sep=8pt] {
  0.1 \& 0.5 \& 0.3 \& 0.7 \\
  0.8 \& 0.2 \& 0.2 \& 0.9 \\
  \vdots \& \vdots \& \vdots \& \vdots \\
  0.6 \& 0.9 \& 0.9 \& 0.1 \\
  0.3 \& 0.4 \& 0.1 \& 0.8 \\
};

\node[above=6pt of X-1-1] {$x_1$};
\node[above=6pt of X-1-2] {$x_2$};
\node[above=6pt of X-1-3] {$x_3$};
\node[above=6pt of X-1-4] {$x_4$};

\node[right=.2cm of X] (arrow) {$\Longrightarrow$};

\matrix (Phi) [matrix of math nodes,
               ampersand replacement=\&,
               left delimiter={[}, right delimiter={]},
               row sep=3pt, column sep=14pt,
               right=.2cm of arrow,
               column 2/.style={nodes={text=orange!90!black}}] {
  1 \& 0 \& 0 \\
  0 \& 1 \& 0 \\
  \vdots \& \vdots \& \vdots \\
  0 \& 0 \& 1 \\
  0 \& 1 \& 0 \\
};

\node[above=6pt of Phi-1-1] {$\phi_1(x)$};
\node[above=6pt of Phi-1-2, text=orange!90!black] {$\phi_2(x)$};
\node[above=6pt of Phi-1-3] {$\phi_3(x)$};

\end{tikzpicture}%
}
\caption{\label{fig:tree-feature-map} Corresponding feature map}
\end{subfigure}

\caption{\label{fig:tree-illustration}Illustration of the feature map implied by decision trees. (A) A single decision tree, with one path highlighted; (b) the induced partition of covariate space; and (c) the corresponding feature map $\phi(x)$. Under a single tree, points in the same partition (leaf of the decision tree) are perfectly similar and points in distinct partitions are perfectly dissimilar. In an ensemble of trees (a forest), the proportion of trees in which two points share a partition defines their similarity, and the corresponding kernel function.}
\end{figure}
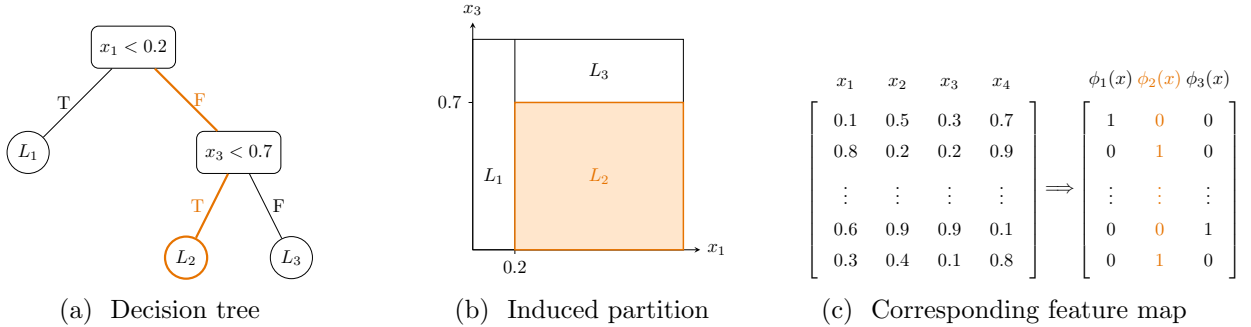

Beginning with a single tree, Figure~\ref{fig:tree-illustration} shows a binary decision tree (a), which implies a partition of covariate space (b). The tree-based features are simply dummy variables for each leaf (c). The implied kernel is the inner product of these:
$$
k(x_i, x_j) = \phi(x_i)^\top\phi(x_j) = I(i,j),
$$
where
\begin{align*}
    I(i,j) = \begin{cases}
1 \text{ if } i \text{ and } j \text{ share a leaf node} \\
0 \text{  otherwise.}
\end{cases}
\end{align*}
Mean-balancing the implied feature map is equivalent to reweighting the control observations to produce a weighted histogram over the partition in Figure \ref{fig:tree-partition} that matches the histogram of treated observations.

A forest is the average over many tree predictions, and accordingly a forest kernel is just the average over many single-tree kernels:
\begin{equation} \label{eqn:forest-kernel}
k(x_i, x_i) = \frac{1}{T}\sum_{t=1}^T k_t(x_i, x_j) = \frac{1}{T}\sum_{t=1}^T I_t(x_i, x_j) = p_{ij},
\end{equation}
where $p_{ij}$ is the proportion of trees in which observations $i$ and $j$ share a leaf. The implied feature map is the concatenation of all the single-tree feature maps scaled by $1/\sqrt{T}$. While features within a tree are orthogonal, even wildly different trees can produce correlated features, especially when the original variables are correlated. So a large tree ensemble will still often produce a kernel with small effective dimension. This motivates the spectral approximations we use below.

The ultimate kernel representation will depend on the size of the tree ensemble, the geometry of the trees in the ensemble, and the process by which they are grown. We consider random forest and BART kernels, which produce tree ensembles in markedly different ways. Both, however, use both outcome and covariate information to construct trees and implied features that are predictive of outcomes.

\paragraph{Random Forest Kernel.}
Random forests are a widely used ensemble learning method that constructs a collection of decision trees (a ``forest''), each trained on a bootstrap sample of the data with random subsets of covariates selected at each split \citep{breiman2001random}. By averaging over many trees, this procedure avoids overfitting and reduces variance, often achieving strong predictive performance \citep{buhlmann2002analyzing,biau2016random,agarwal2023mdi+}. Random forests are popular in part because they are easy to implement, require minimal tuning, and can adapt to relationships in the data that are nonlinear and complex \citep{agarwal2023mdi+}. 
Random forests have been used extensively in causal inference, including as a core estimation component for heterogeneous treatment effects \citep{athey2016recursive,wager2018estimation} and propensity scores \citep{welzgenericml,keele2025balancing}. The implied kernel has the same average-of-many-single-trees form as Equation \eqref{eqn:forest-kernel}:
\begin{align*}
    \bK_{ij}^{\rf} = \bK_{ji}^{\rf} = \frac{1}{T} \sum_{t=1}^T I_t(x_i,x_j).
\end{align*}

The trees in a random forest ensemble are typically grown very deep, resulting in fine partitions. The bootstrap resampling and variable subsampling at each step of the random forest algorithm (along with additional tuning parameters like maximum depth of the trees) limit just how fine the partition can get and reduce correlation across the trees in the ensemble.

\paragraph{BART Kernel.}
Like random forests, BART has proved popular for applied research, with strong empirical performance across a range of tasks, especially when used for causal inference; see \citet{hill2020bayesian} for a comprehensive review. 
Unlike RF, BART is an ensemble of small trees, more similar in spirit to boosted trees than random forests. Unlike boosting, however, the ``Bayesian backfitting'' MCMC algorithm for BART fixes a total number of trees (typically 200) and updates them stochastically in Metropolis-Hastings steps, biasing the trees toward feature representations that best predict outcomes but still randomly exploring a full posterior distribution over tree structures.

One output of the BART MCMC algorithm is a set of posterior draws of tree structures. For $B$ posterior draws we take these as defining our BART kernel via
\begin{align*}
    \bK_{ij}^\bart = \bK_{ji}^\bart = \frac{1}{BT} \sum_{b=1}^B\sum_{t=1}^T I_{bt}(i,j),
\end{align*}
where $I_{bt}(i,j)$ denotes leaf co-membership for $i$ and $j$ in tree $t$ in posterior draw $b$. Again, there is significant redundancy in this representation, both across trees within a posterior sample (since two shallow trees often produce correlated feature representations) and across posterior samples (which generally feature small changes so a subset of trees  from iteration to iteration). This redundancy further motivates using a spectral approximation.

\begin{figure}[tbp]
  \centering

  \begin{subfigure}[c]{0.32\textwidth}
    \centering
    \includegraphics[width=\linewidth]{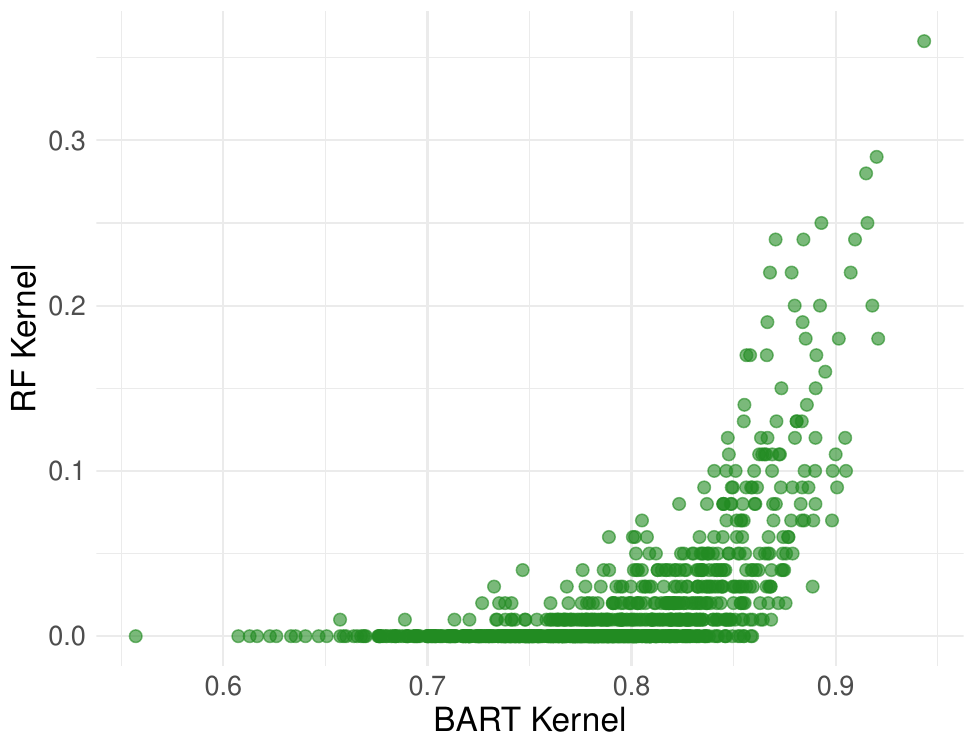}
    \caption{RF vs BART}
    \label{fig:kij-plots:a}
  \end{subfigure}
  \hfill
  \begin{subfigure}[c]{0.32\textwidth}
    \centering
    \includegraphics[width=\linewidth]{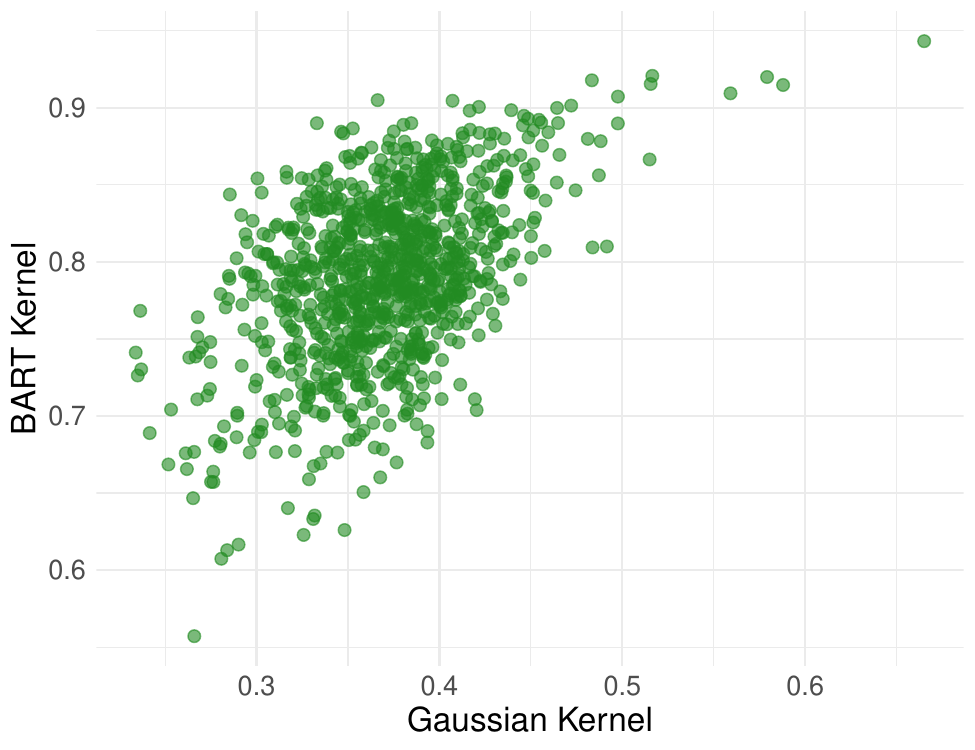}
    \caption{BART vs Gaussian (kbal)}
    \label{fig:kij-plots:b}
  \end{subfigure}
  \hfill
  \begin{subfigure}[c]{0.32\textwidth}
    \centering
    \includegraphics[width=\linewidth]{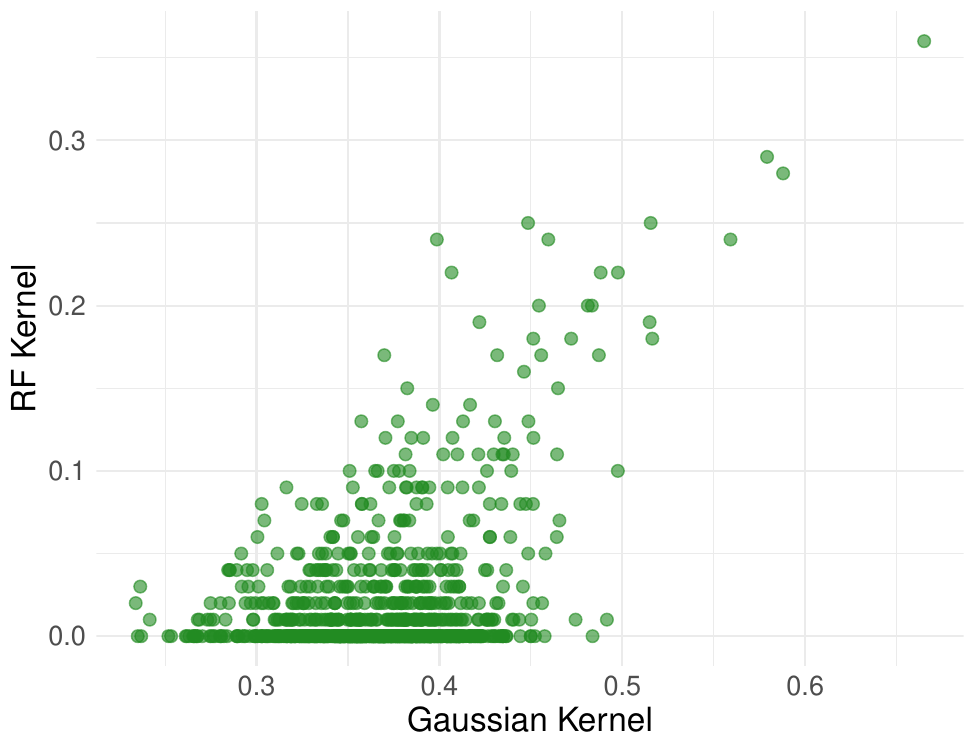}
    \caption{RF vs Gaussian (kbal)}
    \label{fig:kij-plots:c}
  \end{subfigure}

  \caption{Comparison of RF, BART, and Gaussian kernel values. We randomly sample one unit from the observed analysis dataset and plot the corresponding kernel values on each axis.}
  \label{fig:kij-plots}
\end{figure}

\paragraph{Comparison of forest kernels.}
Despite their similar structure, the tree geometries and differences between RF and BART can lead to meaningfully different kernels. Figure \ref{fig:kij-plots} illustrates one example of how these kernels compare (\ref{fig:kij-plots:a}). For comparison, we also illustrate how the forest kernels compare with the Gaussian kernel (\ref{fig:kij-plots:b} and \ref{fig:kij-plots:c}). The figure comes from a single realization of the data-generating process (DGP) outlined in Section \ref{sec:simulation-study} and is shown only to build intuition for geometries induced by these kernels in a given sample. Here we randomly sample one treated unit $i$ from the observed analysis and plot its kernel values $k(i,j)$ where $j \neq i$.

Figure \ref{fig:kij-plots} shows that the all three kernels are generally correlated for this particular draw of target unit $i$. The figure, however, also illustrates key differences in how the two forest kernels encode similarity. 
This instance of RF is based on a smaller number of deeper trees where most pairs are unlikely to co-occur, producing a sparse, localized kernel. 
BART, by contrast, is a sum of many shallow, weak trees with partitions that often overlap. Pairs tend to share at least one terminal region across the ensemble, yielding a smoother, dense kernel.

\subsection{Sample splitting and cross-fitting}
While our forest kernel balancing framework uses the outcome to learn important features, this reliance alters the distinction between the design and analysis stages of causal inference: under the design principle, outcomes are not consulted during the design phase. Using the outcome for both feature construction and treatment effect estimation increases the risk of overfitting and weakens the randomized-trial analogy of observational study design \citep{jin2025cross}. To preserve the design principle while retaining the benefits of outcome-guided kernels, we extend \textit{sample splitting} to our forest kernel balancing approach. We require splitting the data into two samples: a \emph{pilot sample} and an \emph{analysis sample}. 

The pilot sample is used to train the tree-ensemble model. When estimating the ATT, the pilot sample consists solely of control units so that the resulting kernel geometry targets $Y(0)$, the potential outcome we wish to learn. This model is then applied to the analysis sample, which consists of the remaining control units and the treated cohort. Specifically, we use the analysis sample to obtain the kernel matrix, solve for the weights, and estimate the treatment effect. This two-step procedure ensures the outcome is not being reused, similar to the ``honest'' approach of \citet{athey2016recursive} or the ``cross-balancing'' approach of \citet{jin2025cross}. In practice, we perform this sample-splitting or cross-fitting approach multiple times to reduce the variance of the estimates, mimicking repeated cross-validation in prediction tasks. Theoretical properties of this outcome-informed approach are discussed in \citet{jin2025cross}; we refer the reader to this manuscript for further details.

\subsection{Summary: Forest Kernel Balancing} \label{sec:kbal-procedure}

We summarize this section with a procedure for conducting forest kernel balancing. This procedure can be applied multiple times, mimicking repeated cross-fitting.

\begin{algorithm}[H]
\caption{Forest Kernel Balancing}\label{alg:kbal}
\begin{algorithmic}[1]
\State Split the data into a pilot sample (controls only for ATT) and an analysis sample.
\State Using the pilot sample, fit a random forest or BART model of $Y^{\text{pilot}}$ on $\mathbf{X}^{\text{pilot}}$.
\State Apply the fitted model to the analysis sample to obtain the kernel matrix
$\bK^\rf$ or $\bK^\bart$ (denote this $\bK$).
\State Compute the spectral decomposition $\bK = \mathbf{U} \mathbf{D} \mathbf{U}^{\top}$ and form the top-$r$ eigenvectors
scaled by their singular values: $\sigma_kU_k$. Any conventional selection rule can be used to select $r$.
\State If desired, augment the forest kernel with raw covariates and normalize the variance accordingly (see Section~\ref{sec:kbal}); call the resulting covariate matrix $\tilde{\Phi}$.
\State Estimate balancing weights using~\eqref{eqn:bal-wgt-features} with features $\tilde{\Phi}$.
\State Using these weights, compute the estimated ATT via~\eqref{eqn:tauhat}.
\Statex \textit{Optional: repeat the procedure multiple times (cross-fitting) and aggregate the resulting estimates.}
\end{algorithmic}
\end{algorithm}

\section{Simulation Studies} 
\label{sec:simulation-study}
We conduct a simulation study, adapted from previous work on kernel balancing \citep{tarr2025estimating}, to investigate how well our forest kernel balancing methods perform relative to a design-based kernel or to balancing raw covariates alone. 
These simulations show that forest kernel balancing improve both bias and RMSE compared with balancing on raw covariates only or based on design-based kernel balancing.

We defer many implementation details to the supplementary materials, which also include results from a second simulation study based on an alternate DGP \citep{kim2024scalable}.
All balancing weights are estimated using the \texttt{balancer} package in \texttt{R}, which implements the methods outlined in \citet{benmichael2021_lor} and \cite{benmichael2022}.

\subsection{Simulation Setup}
Our primary simulation DGP is from \citet{tarr2025estimating}, which in turn was adapted from \citet{wong2018kernel}; we refer to these papers for further discussion. 
In this simulation, both the propensity score and outcome models are nonlinear in the observed covariates and contain multiple interactions and higher order covariate transformations. 
Let $X  = (X_{1},\cdots, X_{10})^{\top}$ be 10 covariates constructed to be highly nonlinear: for $W_1, \ldots, W_{10}$ iid standard Gaussians, $X_1 = \exp(W_1/2)$, $X_2 = W_2/(1 + \exp(W_1))$, $X_3 = (W_1W_3/25 + 0.6)^3$, $X_4 = (W_2 + W_4 + 20)^2$ and $X_j = W_j$ ($j = 5,...,10$). The propensity score is logistic with $P(Z=1 \mid W) = \text{logit}^{-1}(-W_{1} - 0.1W_{4})$
 Finally, the outcome model is $Y(Z) = 200 + 10Z + (1.5Z - 0.5)(27.4 W_{1} + 13.7 W_{2} + 13.7 W_{3} + 13.7 W_{4}) + \varepsilon$, where $\varepsilon \sim \mathcal{N}(0,1)$. The target estimand is the ATT defined in Equation \eqref{eqn:att}. 

We balance on the three different feature groupings introduced in Section \ref{sec:forest-kernels}: raw covariates only, kernel features only, and a combination of both. For the latter two feature groupings, we balance on a varying number of principal components to assess the sensitivity of our results to this choice. We use 1000 Monte Carlo replications and set the analysis sample size to 1000. For the pilot sample, we draw an additional sample of 1000 units and keep only the control units, resulting in roughly 500 pilot units per simulation run. We discard the treated units because our target estimand is the ATT. The random forest/BART model is trained on the pilot sample using 100 trees and the kernel is constructed by fitting this trained model to all units in the analysis sample. We measure \textit{relative} absolute bias, defined as the absolute value of the average difference between the true treatment effect and the estimated treatment effect, divided by the true treatment effect, 
$\left\lvert (\hat{\tau}- \tau) / \tau\right\rvert$. 
We record relative absolute bias since the direction and magnitude of the bias differs for each simulation run. We also measure the \textit{relative} root mean-squared error (RMSE),
$
    \sqrt{(\hat{\tau} - \tau)^2/\tau},
$
which we report in the main text.
Results for the relative absolute bias are similar in trend and magnitude and can be found in Figure \ref{fig:sim1-relbias} in the supplement.

\subsection{Simulation Results} 
\label{sec:simulation-results}
\begin{figure}[ht]
    \centering
    \includegraphics[width=0.95\linewidth]{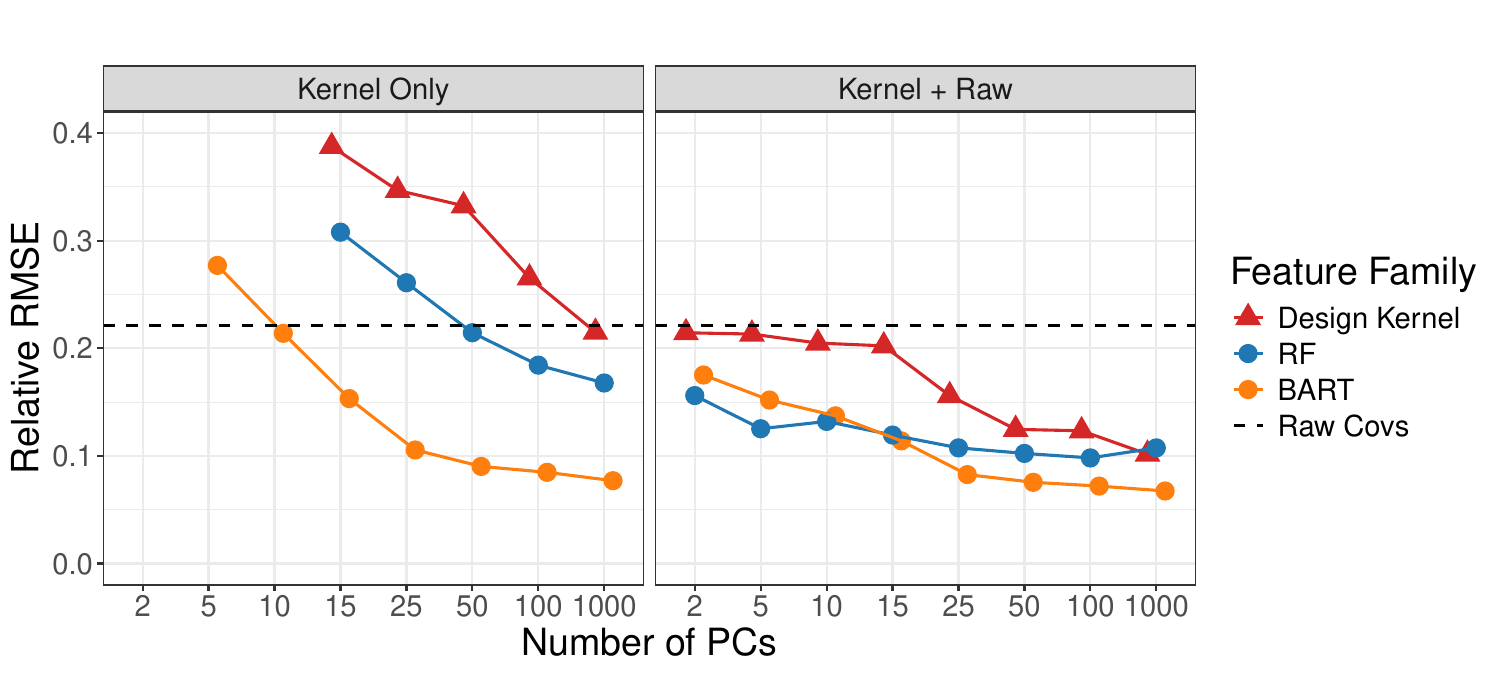}
    \caption{Absolute relative RMSE for design-based and forest kernels, varying the number of principal components to balance on. The left panel shows results using only kernel features (principal components); the right panel includes both raw features and kernel features. The black dashed line indicates performance from using raw covariates alone. Full results, including points outside the axis limits, are reported in Supplementary Table \ref{tab:sim1-full}.}
    \label{fig:sim1-relrmse}
\end{figure}

The relative RMSE is shown in Figure \ref{fig:sim1-relrmse}, plotted across a varying number of principal components (horizontal axis). An analogous plot for the relative bias can be found in Figure \ref{fig:sim1-relbias} in the supplement.
The solid lines represent different feature representations: blue represents the random forest kernel, orange represents the BART kernel, and red represents the Gaussian design-based kernel. Overall, the three kernel balancing methods exhibit similar trends in bias and RMSE across principal components. Balancing only on kernel features, however, generally performs worse than using raw covariates, especially with too few principal components. This likely reflects the fact that such a low-dimensional approximation of the kernel cannot sufficiently capture all sources of confounding. In particular, since kernel balancing emphasizes more complex interactions, this feature representation fails to account for the linear basis explained by the raw covariates. 

Balancing on both the kernel features and raw covariates results in much better performance, with forest kernels outperforming the raw covariates and design-based kernel by nearly twofold for moderate numbers of included principal components. 
Adding more principal components continues to improve RMSE, although the gains eventually plateau for the BART and RF kernels, especially when these methods also balance the raw covariates.
Across both feature groupings, the forest kernels strictly outperform the design-based kernel, demonstrating that the outcome model produces features that better capture confounding mechanisms.

In addition to evaluating estimation accuracy, we also assess the quality of the weights. In particular, we examine \textit{the effective sample size (ESS)}:
\begin{align}
\label{eqn:ess}
    \text{ESS} = \frac{(\sum_{Z_i = 0} w_i)^2}{\sum_{Z_i = 0} w_i^2},
\end{align}
which measures loss of precision due to weight variability. In our simulation study, the ESS shows a similar pattern where the value eventually degrades when balancing on a large basis (see Figure \ref{fig:sim1-ess} in the supplement). These metrics suggest that our kernel balancing methods are generally robust to the number of principal components, provided the dimensionality is not pushed to extremes. 

We also replicate both simulations using logistic regression to estimate the propensity score (and corresponding weights) instead of balancing weights. The results can be found in Figures \ref{fig:sim1-relbias-ipw} and \ref{fig:sim1-relrmse-ipw} of the supplement. The substantive conclusions remain largely unchanged, indicating that our findings are not sensitive to the choice of weighting estimator.

To assess performance in higher dimensions, we implemented an additional simulation where we varied the number of covariates $q \in \{10, 50, 100\}$ and the sample size $n \in \{1000, 2000\}$. This extension of the data-generating process ensures that blocks of covariates carry the same confounding structure; see Section \ref{sec:varying-q-n} in the supplement for full details. The forest kernels continue to outperform the design-based kernel for moderate spectral truncations. As $q$ grows relative to $n$, the design-based kernel no longer dominate the raw covariates alone, while the forest kernels retain lower relative RMSE.

\begin{revision}
We also assess whether these improvements depend on our choice of weighting procedure by replacing the balancing weights with covariate balancing propensity score weighting (CBPS) \citep{imai2014covariate}. We supply the same feature representations to CBPS in the simulation varying the covariate dimension. The learned forest features continue to improve on CBPS applied to the raw covariates alone, so the gains from outcome-guided feature learning do not depend on the weighting procedure. Unlike our main implementation, however, CBPS degrades when the number of balanced components is large relative to the sample size. The full comparison is shown in Appendix \ref{sec:cbps}.
\end{revision}



\section{Empirical Applications} \label{sec:application}
We evaluate our forest kernel balancing methods using two empirical applications.  The first application is a within-study comparison from \citet{keller2025new} that explores the relative effectiveness of a mathematics tutoring module.
The second application, from \citet{blattman2010consequences} and re-analyzed in \citet{wong2018kernel}, is an observational study of child soldiering in Uganda.

\subsection{Within-Study Comparison} \label{sec:wsc}

\subsubsection{Background}

We first evaluate our forest kernel balancing methods using data from \citet{keller2025new}. This within-study comparison (WSC) design \citep{wong2018can} allows us to directly compare estimated effects from an observational study with an experimental benchmark. 

The goal of the study is to assess the impact of an instructional training module that focused on either math or vocabulary instruction. 
Participants (\(n = 2200\)) first state a preference for math training, denoted $X_{\text{math}} = 1$, or vocabulary training, $X_{\text{math}} = 0$. Participants are then randomly assigned to math training, $Z_{\text{math}} = 1$, or vocabulary training, $Z_{\text{math}} = 0$, with randomization stratified by $X_{\text{math}}$. This creates four groups: \(\{X_{\text{math}}=1, Z_{\text{math}}=1\}\) preferred math and were assigned math ($n=288$); \(\{X_{\text{math}}=1, Z_{\text{math}}=0\}\) preferred math and were assigned vocabulary ($n=319$); \(\{X_{\text{math}}=0, Z_{\text{math}}=1\}\) preferred vocabulary and were assigned math ($n=776$); \(\{X_{\text{math}}=0, Z_{\text{math}}=0\}\) preferred vocabulary and were assigned vocabulary ($n=817$). 
While the original study included many analyses, we focus here on the impact on the math posttest, which we denote $Y$. We also have access to 25 baseline covariates (in addition to training preference), which we denote $X$.

One important feature of a within-study comparison design is that, because treatment assignment is randomized within each preference stratum, the contrast between groups 1 and 2 (both prefer math, randomly assigned to math or vocabulary) provides an experimental benchmark for the effect of math instruction on those who would prefer math if offered the choice: $E[Y(Z_{\text{math}} = 1) - Y(Z_{\text{math}} = 0) \mid X_{\text{math}} = 1]$. Using this randomized contrast, \citet{keller2025new} found that participating in mathematics tutoring resulted in slightly higher mathematics posttest scores (0.79, SE = 0.28). 
Because groups 3 and 4 are randomized within the vocabulary preference stratum, they are not comparable to groups 1 and 2, creating the observational cohort. The key question is whether we can replicate this experimental benchmark by instead comparing the first and fourth groups, $\{X_{\text{math}} = 1, Z_{\text{math}} = 1\}$ vs $\{X_{\text{math}} = 0, Z_{\text{math}} = 0\}$, and adjusting for the 25 baseline covariates $X$ (excluding $X_{\text{math}}$).


Before applying any transformations, we examine results using the original covariates as a baseline. All methods produce estimates close to one another and consistent with simple adjustment, with confidence intervals overlapping the experimental benchmark (Supplementary Figures \ref{fig:wsc-full-log} and \ref{fig:wsc-full-exp}, untransformed panels). The linear basis sufficiently captures the relevant confounding in this setting, so the kernel-based methods do not differ appreciably from raw adjustment. We note that this is not a shortcoming of forest kernel balancing; there is simply no additional structure for the kernel methods to recover when the covariate-outcome relationships are well-captured by the raw covariates themselves. 

Therefore, to examine performance in a regime where the linear basis is no longer adequate, we follow a strategy from \citet{hazlett2020kernel} and apply a logarithmic transformation ($x \mapsto \log(x + 1)$) to each continuous covariate before performing the analysis. 
As discussed in Section \ref{sec:methodology}, forest kernels are largely invariant to such marginal transformations, whereas the linear and Gaussian kernels are scale-sensitive.
The transformation distorts the linear basis without changing the underlying causal structure: the same confounding mechanisms remain, but the covariates are no longer on a scale where standard covariate adjustment captures them. This transformation assesses whether the forest kernels can recover the experimental benchmark in a regime where competing methods cannot. Finally, all our estimates average over ten cross-fit iterations, as in our main proposal above.

\subsubsection{Results}
\begin{figure}[tb]
    \centering
    \includegraphics[width=0.5\linewidth]{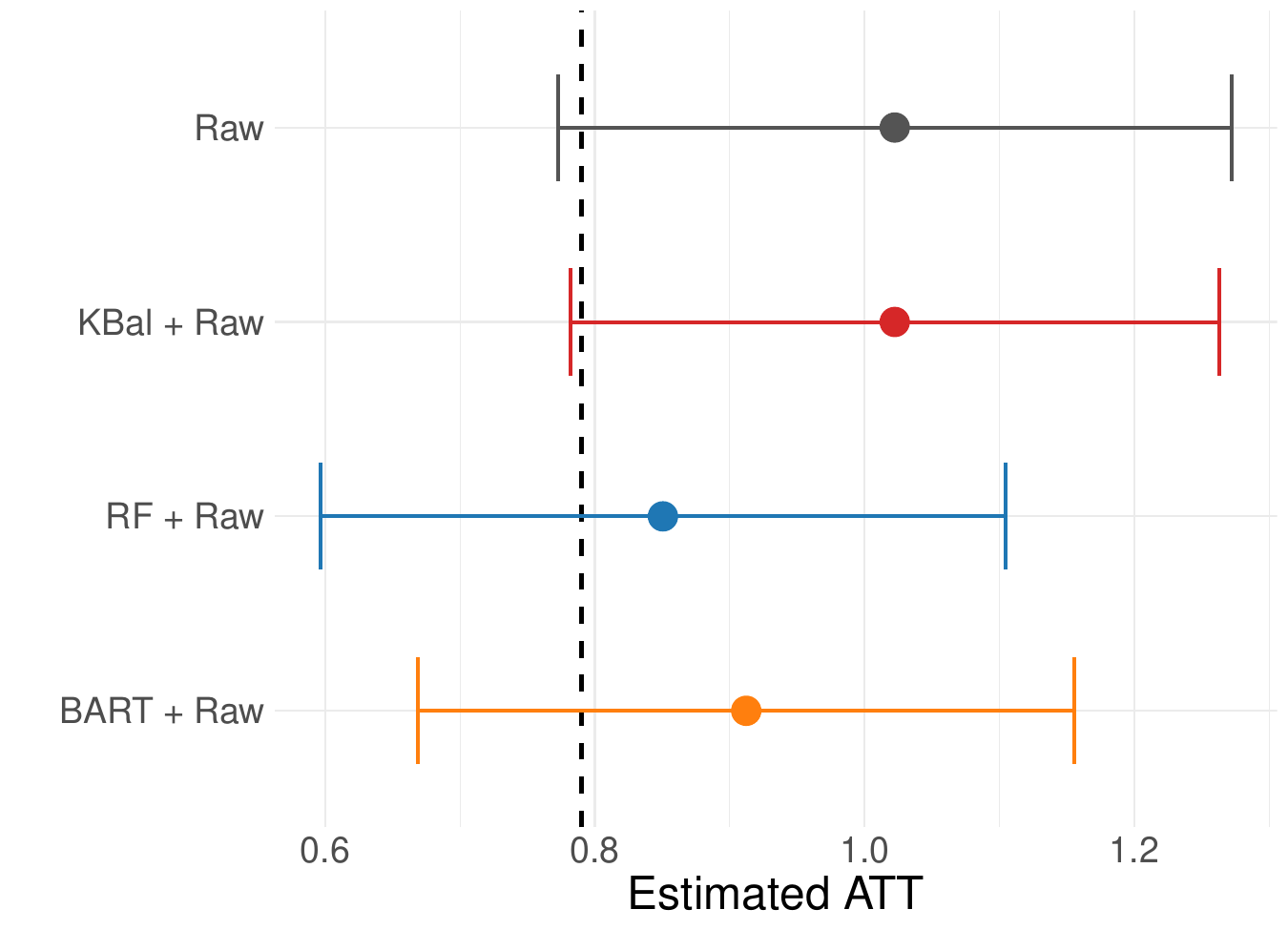}
    \caption{WSC RCT benchmarking results. The black dashed line indicates the experimental benchmark and the colored point estimates are estimated with observational data with different adjustment methods. Error bars denote 95\% confidence intervals. ESS values for each feature representation are as follows: Raw: 210.9; KBal + Raw: 216; BART + Raw: 103.7; RF + Raw: 94.1 (BART and RF estimated via cross-fitting, which reduces ESS).}
    \label{fig:wsc-main}
\end{figure}

Figure \ref{fig:wsc-main} shows the results from the data analysis. The experimental benchmark is indicated by the black dashed line; each estimate represents a different feature representation for balancing weights. 
The gray estimate shows raw covariates only, the red estimate applies the design-based Gaussian kernel \citep{hazlett2020kernel}, and blue and orange apply forest kernel balancing using random forest and BART, respectively. After inspecting a scree plot of principal components (see supplementary Figure \ref{fig:scree-wsc}), we choose to balance on 5 principal components for illustration only (in addition to the 25 baseline covariates themselves); we find that the results are largely insensitive to the number of principal components in this example.
A complete set of results across multiple principal components is reported in supplementary Figure \ref{fig:wsc-full-log}. For our analyses, we do not use cross-fitting to estimate the raw covariate and Gaussian kernel estimates, although this can be implemented if desired by the analyst.

The estimates from the transformed raw covariates and transformed Gaussian kernel deviate substantially from the experimental benchmark, with confidence intervals that nearly exclude it. In contrast, the forest kernel estimates are much closer to the experimental benchmark, indicating the value of using an outcome-guided approach to identify prognostic features.

\subsection{Application: Forced Youth Military Service} \label{sec:soldiering}

We consider the \citet{blattman2010consequences} study of forced youth military service in Uganda, re-analyzed by \citet{wong2018kernel} in their seminal kernel balancing paper. In the 1980s and 1990s, there was a longstanding conflict in Northern Uganda with the Lord's Resistance Army (LRA), a rebel group led by Joseph Kony \citep{doom1999kony}. During this time, the LRA sustained its militia largely through indiscriminate mass abductions rather than voluntary recruitment. In a remarkable data collection effort, the 2005–2006 Survey of War Affected Youth, \citet{blattman2010consequences} surveyed 741 males who were at risk of abduction during this conflict; of these, 462 had been previously abducted by the LRA. The authors then argue that abduction is a plausibly (conditionally) exogenous exposure:
abductions occurred indiscriminately within villages and only youth within a certain age range were abducted for military service, making location and age key covariates for adjustment. 
Other covariates include the youth's orphan status, household size, parental education, and amount of livestock owned. 

Following \citet{blattman2010consequences}, we focus on educational attainment as the primary outcome, as measured by years of schooling. 
We obtain the forest kernel by averaging over ten cross-fit estimates. For each, we partition the control group ($n = 279$) into two approximately equal halves, using one half for model fitting and the other for analysis, we then swap the roles. 
As we discuss above, a key hyperparameter is choosing the rank of the kernel approximation to balance. Following our previous example, we examine a scree plot of principal components (see Figure \ref{fig:scree-soldiering} in the supplement) and choose to balance on 5 principal components in addition to the 15 baseline covariates themselves.

\subsubsection{Data Analysis}

\begin{figure}[tb]
    \centering
    \includegraphics[width=0.65\linewidth]{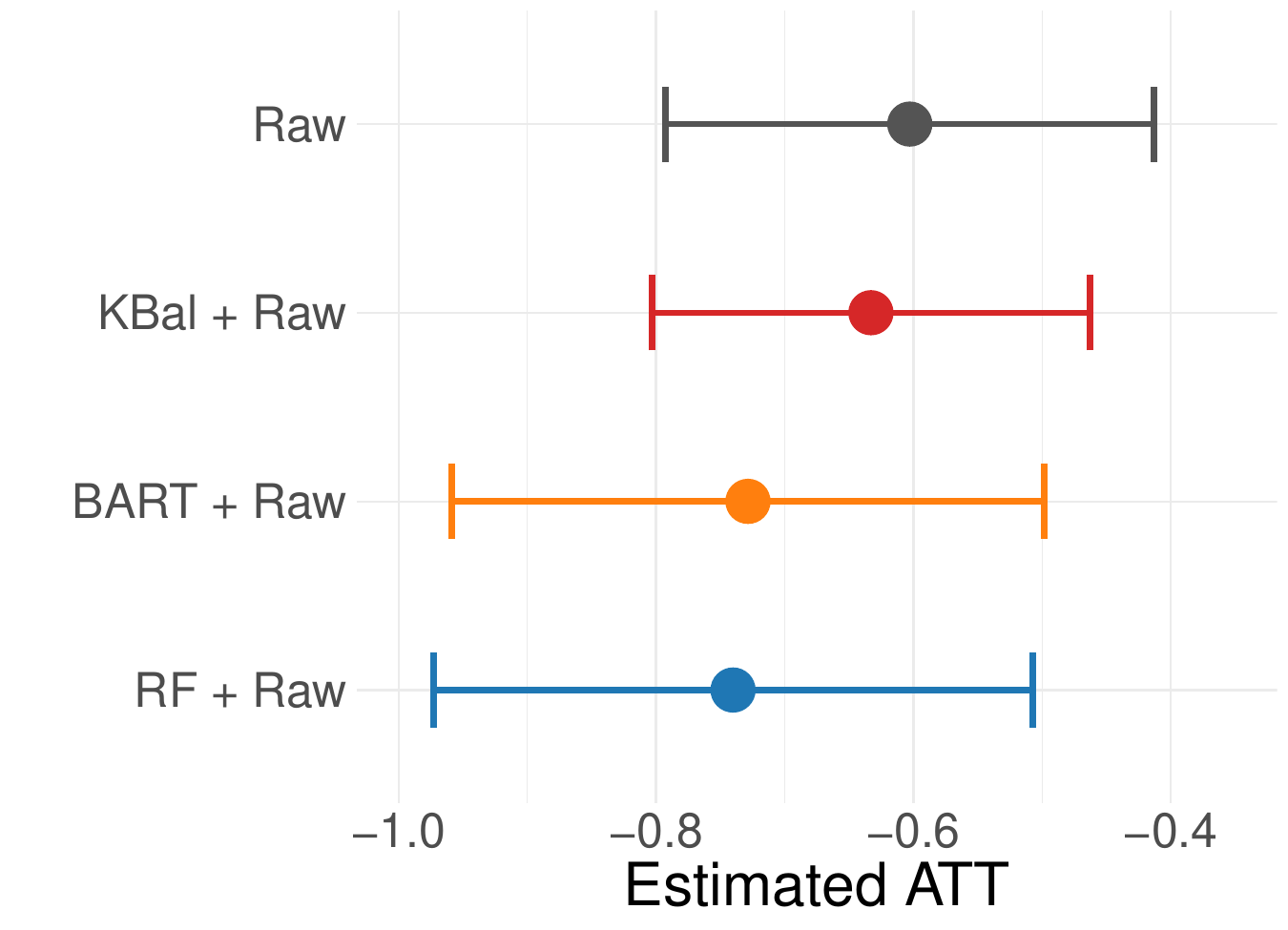}
    \caption{Child soldiering point estimates for 5 principal components. Error bars denote 95\% confidence intervals. ESS values for each feature representation are as follows: Raw: 133; KBal + Raw: 171; BART + Raw: 80; RF + Raw: 78 (BART and RF estimated via cross-fitting, which reduces ESS).}
    \label{fig:soldiering-main}
\end{figure}

Figure \ref{fig:soldiering-main} shows the results from the data analysis. Balancing the raw covariates alone and balancing both the raw covariates and the low-rank approximation to the Gaussian design-based kernel yields estimated effects of -0.6 and -0.63 years of education, respectively. By contrast, the forest kernel estimates are larger, at -0.73 and -0.74 years for BART and RF, respectively, albeit with wider confidence intervals and smaller effective sample sizes due to cross-fitting.\footnote{The effective sample size for cross-fit Gaussian kernel balancing weights is 76, comparable to those for BART and RF.} Again, we do not use cross-fitting to estimate the raw covariate and Gaussian kernel estimates.

\begin{figure}[tb]
    \centering
    \includegraphics[width=0.8\linewidth]{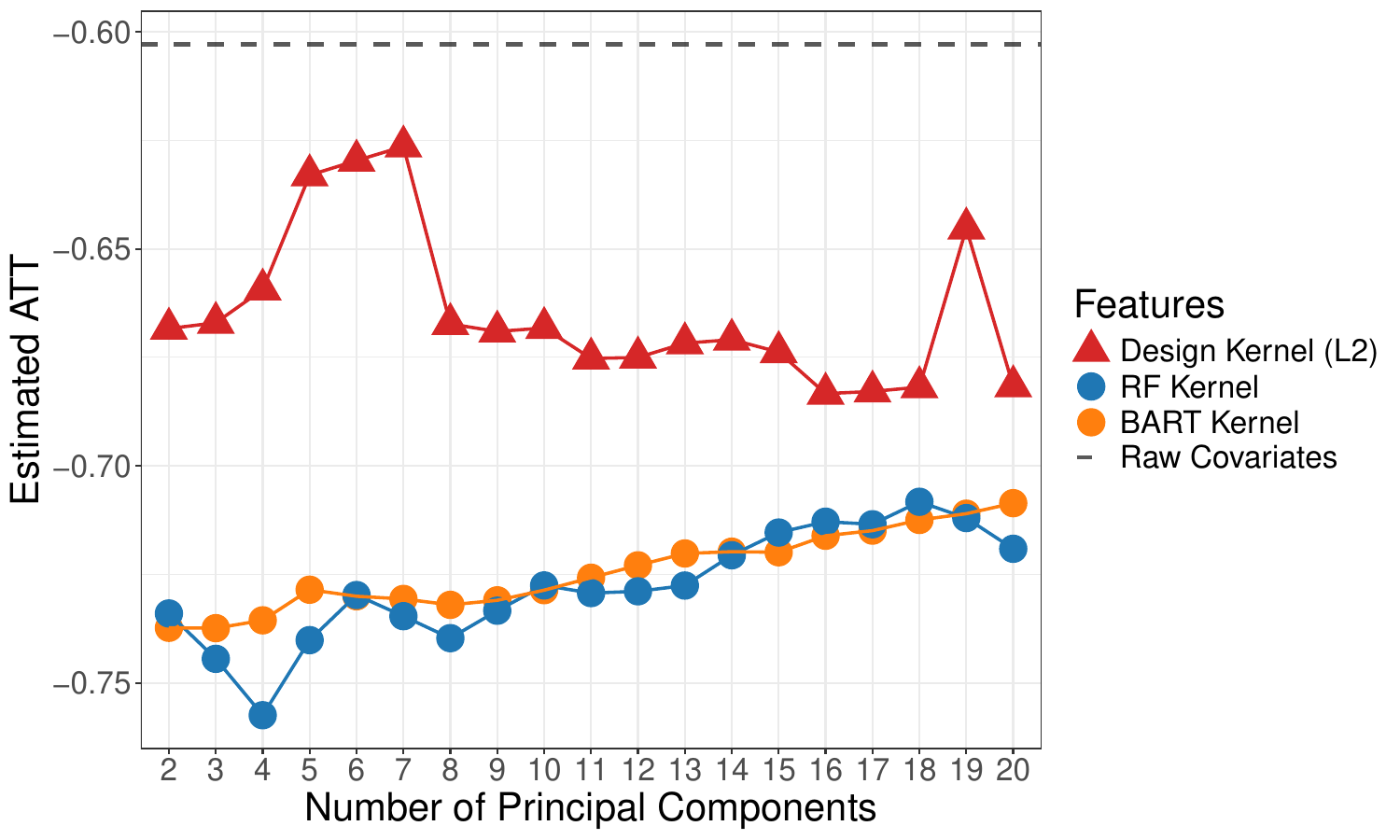}
    \caption{Estimated ATT for child soldiering across varying numbers of included principal components. These estimates balance on both raw features and kernel features. The black dashed line is the estimate from balancing using raw covariates alone.}
    \label{fig:soldiering-full}
\end{figure}

While there is no ground truth in this application, we can assess the stability of each method to changes in hyperparameter choice, following the ``stability'' principle of veridical data science \citep{yu2020veridical,yu2024veridical,rewolinski2025pcs}.
Figure \ref{fig:soldiering-full} plots the ATT estimates across a varying number of principal components, similar to the simulation study. As the number of principal components increases, the forest kernel estimates in the right-hand plot increase slightly but remain close to one another. This stability suggests that, while the RF and BART kernels have different geometries, the weights produced by each adjust for similar mechanisms of confounding. 
By contrast, the design-based Gaussian kernel is more sensitive to the number of principal components: the estimates fluctuate more noticeably as additional components are included. This behavior likely reflects the outcome-agnostic nature of design-based kernels, which encodes similarity solely on the basis of covariates rather than outcome relevance. As a result, adding more components may capture spurious variation which may or may not capture outcome-relevant confounding. The greater stability of the forest kernels suggests that incorporating outcome information helps distinguish between outcome-relevant features and those that merely reflect random covariate variation. 

\section{Discussion} \label{sec:conclusion}
Creating comparable groups is a key tenet of a successful observational study. Balancing weight methods seek to create such groups by producing weights that balance covariate distributions across groups. The challenge lies in correctly accounting for sources of confounding beyond raw covariates, as there exist many complex but important covariate-outcome relationships that may confound a study. In this paper, we combine principles from kernel balancing and outcome models to create an outcome-guided kernel balancing method. Specifically, we leverage the fact that random forest and BART machine learning models implicitly estimate a kernel matrix based on leaf node co-occurrences. These forest kernel matrices reflect complex relationships through the lens of the model, serving as additional features to balance on in weighting. We demonstrate how balancing on a combination of raw and kernel-based features leads to more accurate effect estimation both in simulation and in empirical studies.

There are many promising areas of future work. One is to integrate other aspects of observational studies to the kernel balancing framework, particularly sensitivity analysis for unobserved confounding. Most sensitivity frameworks assume a linear additive basis and posit how an unobserved raw covariate may alter an observational result. Yet, as suggested by our work, the strength of unobserved confounding may also arise from covariate misspecification. Integrating feature representations with sensitivity analysis could help capture this dimension, providing a clearer view of potential threats to observational conclusions. A rich body of sensitivity analyses for weighted estimators already exists \citep{zhao2019sensitivity,soriano2023interpretable,huang2025variance,shen2025calibrated}, so extending these frameworks to kernel balancing offers a promising line of research.
In independent work, \citet{de2025data} propose a multivariate random forest approach that jointly models treatment and outcome, enabling a richer representation of confounding structure than random forests that predict the outcome alone. This approach is especially promising for minimizing adjustment on instrumental variables or on overly noisy covariates. Future iterations of our work could incorporate this multivariate splitting criterion for random forests, as well as analogous extensions to BART with multivariate outcomes \citep{um2023bayesian}.
Finally, our work occupies a middle ground between outcome modeling and weighting in observational studies. Another future direction is to further investigate the role of this intermediate approach as part of a broader doubly robust/augmented weighting approach \citep{chernozhukov2018dml,bruns2025augmented}.

\section*{Acknowledgments}
The authors thank Gary Chan and Raymond Wong for their assistance in obtaining the child soldiering data.
The authors also thank Chad Hazlett for helpful comments.
A.F., L.K., and E.B-M. were supported in part by the Institute of Education Sciences (IES), U.S. Department of Education, through Grant No. R305D240036.
A.S. is partially supported by the National Science Foundation (NSF) Graduate Research Fellowship under Grant No. 2146752. Any opinion, findings, and conclusions or recommendations expressed in this material are those of the authors and do not necessarily reflect the views of the IES or the NSF.

\clearpage
\singlespacing
\bibliographystyle{chicago}
\bibliography{ob-weight}

\clearpage
\doublespacing
\setcounter{page}{1}
\begin{center}
    \singlespacing
    \Large
    \textbf{Supplementary Materials:} \\Forest Kernel Balancing Weights: Outcome-Guided Features for Causal Inference \\
\end{center}

\appendix

\counterwithin{figure}{section} 

\counterwithin{table}{section} 

\section{Additional figures}
\begin{figure}[htbp]
  \centering
  \begin{subfigure}[t]{0.48\textwidth}
    \centering
    \includegraphics[width=\linewidth]{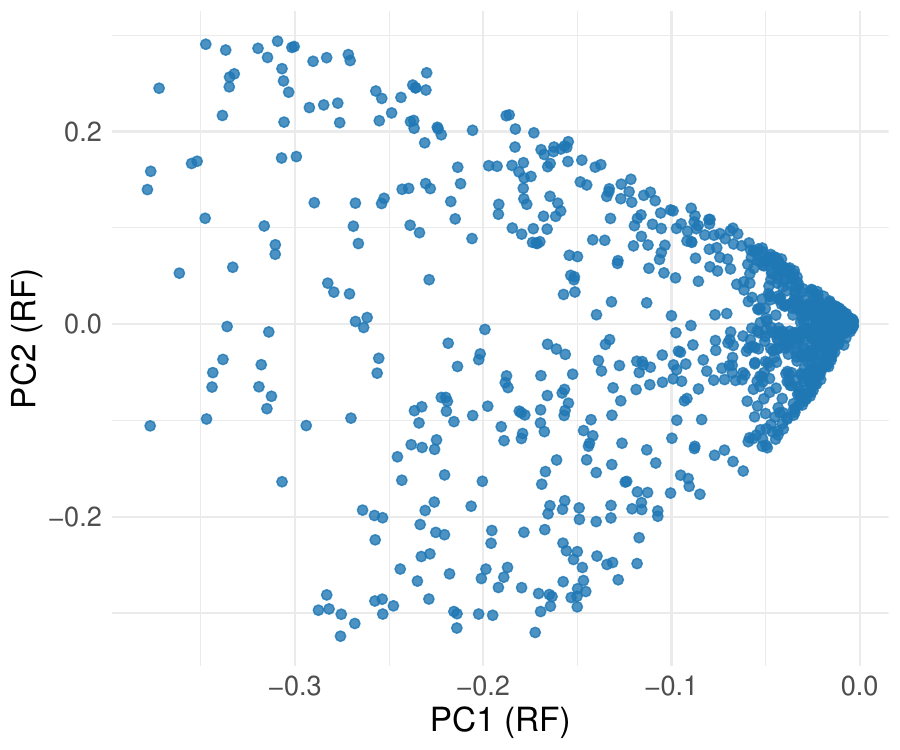}
    \subcaption{Random forest principal components}
    \label{fig:rf_pc}
  \end{subfigure}\hfill
  \begin{subfigure}[t]{0.48\textwidth}
    \centering
    \includegraphics[width=\linewidth]{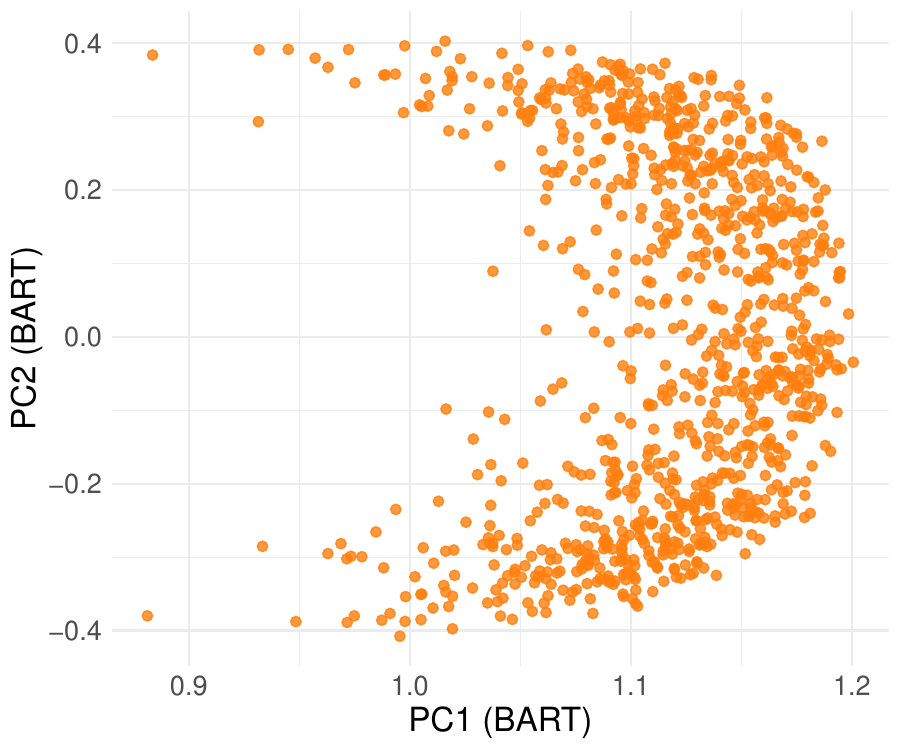}
    \subcaption{BART principal components}
    \label{fig:bart_pc}
  \end{subfigure}
  \caption{Side-by-side comparison of random forest (left) and BART (right) principal-component scatterplots.}
  \label{fig:rf_bart_side_by_side}
\end{figure}

Figure \ref{fig:rf_bart_side_by_side} provides a simple illustration of the first two principal components (PCs) of the RF (\ref{fig:rf_pc}) and BART (\ref{fig:bart_pc}) kernels. The plots come from the same realization of the DGP used to present Figure 2 in the main text.
We note that different DGPs (and even different draws of the same DGP) can produce qualitatively different PC configurations.

\section{Additional results from Simulation 1} \label{sec:extra-simulation-results}
In this section, we present additional results from Simulation 1 (in the main text) that were omitted for clarity.

\subsection{Full bias and RMSE results}
\begin{figure}[ht]
    \centering
    \includegraphics[width=0.85\linewidth]{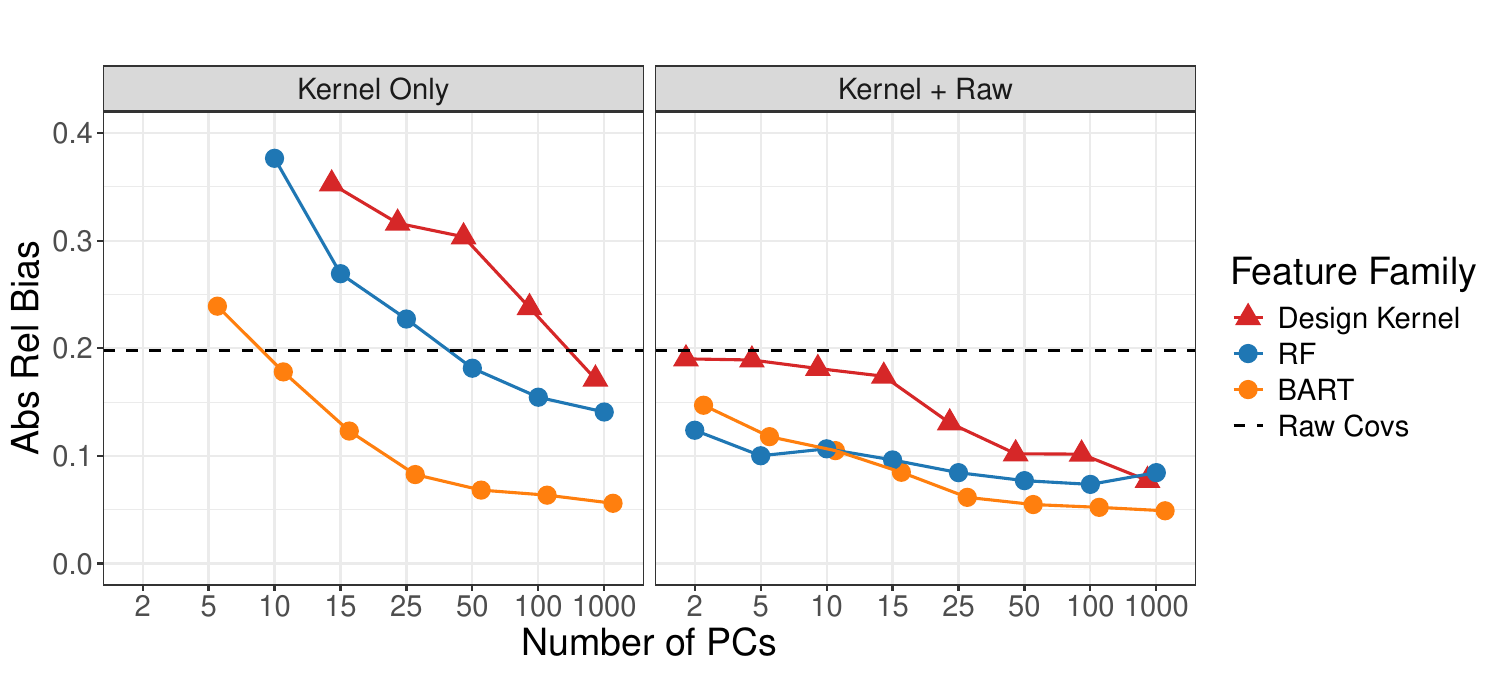}
    \caption{Absolute relative bias for design-based and forest kernels, varying the number of principal components to balance on. The left panel shows results using only kernel features (principal components); the right panel includes both raw features and kernel features. The black dashed line indicates performance from using raw covariates alone. Full results, including points outside the axis limits, are reported in Supplementary Table \ref{tab:sim1-full}.}
    \label{fig:sim1-relbias}
\end{figure}

Figure \ref{fig:sim1-relbias} shows the relative bias from this simulation, analogous to Figure \ref{fig:sim1-relrmse} in the main text.

Table \ref{tab:sim1-full} shows the results from the figure in Section \ref{sec:simulation-study} the main text. As certain points were omitted from the plot due to large bias/RMSE magnitudes, this table is presented for completeness.

\begin{table}[ht]
\centering
\begin{tabular}{cccc}
  \hline
\textbf{Feature Grouping} & \textbf{Number of PCs} & \textbf{Absolute Relative Bias} & \textbf{Relative RMSE} \\ 
  \hline
bart\_only & 2.00 & 0.61 & 0.99 \\ 
  bart\_only & 5.00 & 0.23 & 0.27 \\ 
  bart\_only & 10.00 & 0.17 & 0.22 \\ 
  bart\_only & 15.00 & 0.12 & 0.15 \\ 
  bart\_only & 25.00 & 0.08 & 0.10 \\ 
  bart\_only & 50.00 & 0.07 & 0.09 \\ 
  bart\_only & 100.00 & 0.06 & 0.08 \\ 
  bart\_plus & 2.00 & 0.14 & 0.17 \\ 
  bart\_plus & 5.00 & 0.11 & 0.15 \\ 
  bart\_plus & 10.00 & 0.10 & 0.13 \\ 
  bart\_plus & 15.00 & 0.08 & 0.10 \\ 
  bart\_plus & 25.00 & 0.06 & 0.08 \\ 
  bart\_plus & 50.00 & 0.05 & 0.07 \\ 
  bart\_plus & 100.00 & 0.05 & 0.07 \\ 
  kbal\_only & 2.00 & 1.63 & 1.79 \\ 
  kbal\_only & 5.00 & 1.59 & 1.74 \\ 
  kbal\_only & 10.00 & 0.40 & 0.46 \\ 
  kbal\_only & 15.00 & 0.35 & 0.40 \\ 
  kbal\_only & 25.00 & 0.31 & 0.36 \\ 
  kbal\_only & 50.00 & 0.30 & 0.35 \\ 
  kbal\_only & 100.00 & 0.23 & 0.27 \\ 
  kbal\_plus & 2.00 & 0.19 & 0.22 \\ 
  kbal\_plus & 5.00 & 0.19 & 0.22 \\ 
  kbal\_plus & 10.00 & 0.18 & 0.21 \\ 
  kbal\_plus & 15.00 & 0.17 & 0.21 \\ 
  kbal\_plus & 25.00 & 0.13 & 0.16 \\ 
  kbal\_plus & 50.00 & 0.10 & 0.13 \\ 
  kbal\_plus & 100.00 & 0.10 & 0.13 \\ 
  raw & 0.00 & 0.20 & 0.23 \\ 
  rf\_only & 2.00 & 0.52 & 0.57 \\ 
  rf\_only & 5.00 & 0.48 & 0.53 \\ 
  rf\_only & 10.00 & 0.37 & 0.42 \\ 
  rf\_only & 15.00 & 0.25 & 0.29 \\ 
  rf\_only & 25.00 & 0.22 & 0.25 \\ 
  rf\_only & 50.00 & 0.17 & 0.20 \\ 
  rf\_only & 100.00 & 0.15 & 0.18 \\ 
  rf\_plus & 2.00 & 0.12 & 0.15 \\ 
  rf\_plus & 5.00 & 0.10 & 0.12 \\ 
  rf\_plus & 10.00 & 0.10 & 0.13 \\ 
  rf\_plus & 15.00 & 0.09 & 0.12 \\ 
  rf\_plus & 25.00 & 0.08 & 0.10 \\ 
  rf\_plus & 50.00 & 0.07 & 0.09 \\ 
  rf\_plus & 100.00 & 0.07 & 0.09 \\ 
   \hline
\end{tabular}
\caption{Complete balancing weight results from simulation 1. Feature representations with \_only are kernel only features, \_plus are kernel + raw covariates.} 
\label{tab:sim1-full}
\end{table}

\clearpage

\subsection{Varying covariate dimension} \label{sec:varying-q-n}
In this section, we modify the existing simulation from the main manuscript to investigate the performance of forest kernel balancing weights in high dimensions. In particular, we generalize the DGP from the simulation in Section \ref{sec:simulation-study} to an arbitrary dimension $q$. The covariates are built from $q$ independent standard Gaussian latent variables grouped into $B$ consecutive blocks of ten. Within each block, we apply the same nonlinear transformations as the original process: the first four covariates are nonlinear and confounding and the remaining six are noise, holding the fraction of confounding covariates constant at 40\% across all dimensions. Each block contributes the same terms to the propensity and outcome models. Because the blocks are independent, summing their contributions would inflate the variance of the propensity and outcome linear predictors as $q$ grows, which would degrade overlap and change the signal-to-noise ratio. To isolate the effect of dimension rather than confounding strength, we rescale the accumulated propensity and outcome signals by $\sqrt{B}$, where $B$ is the number of blocks. This holds the variance of both linear predictors approximately constant across $q$, so that propensity overlap and outcome signal-to-noise remain comparable. 

We consider $q \in \{10, 50, 100\}$ and $n \in \{1000, 2000\}$, where $n$ is the analysis sample size (note that $q=10$ and $n=1000$ recovers the simulation setting in the main text). As in the main text, the pilot sample is drawn separately and reduced to its control units, the forest model is trained on the pilot sample, and the kernel is constructed by applying the trained model to the analysis sample. Diagnostics are reported over 1000 Monte Carlo replications, balancing on both raw covariates and the kernel features (the ``Kernel + Raw'' representation that performed best in Section \ref{sec:simulation-study}). 

\begin{figure}[ht]
    \centering
    \includegraphics[width=0.85\linewidth]{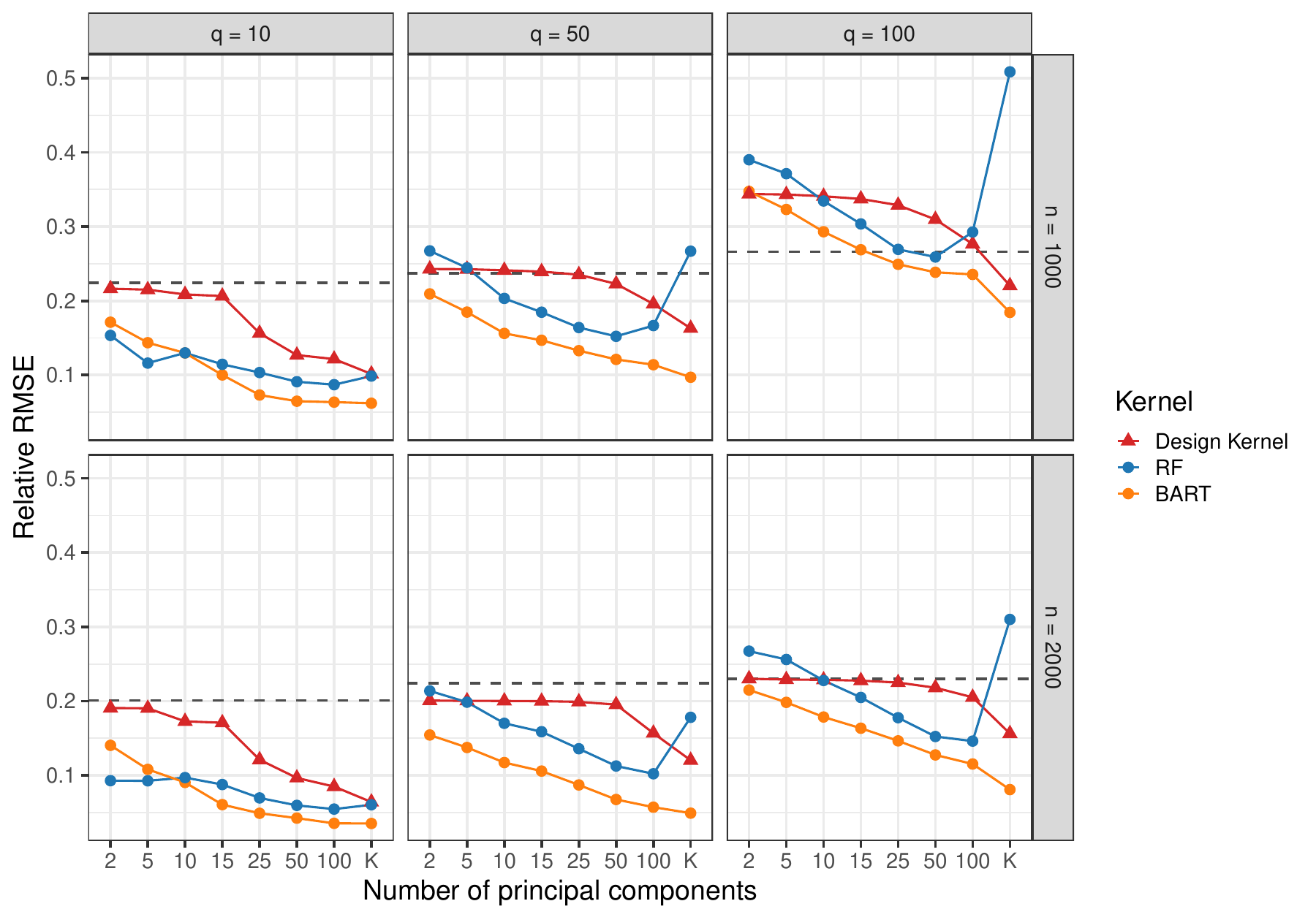}
    \caption{Relative RMSE for design-based and forest kernels, varying the covariate dimension $q$ and the analysis sample size $n$. Columns correspond to the number of covariates ($q = 10, 50, 100$) and rows to the analysis sample size ($n = 1000, 2000$). All estimates balance on both raw covariates and kernel features, varying the number of principal components on the horizontal axis, where $K$ denotes the full kernel. The black dashed line indicates performance from using raw covariates alone.}
    \label{fig:sim1-revision-rmse}
\end{figure}

\begin{figure}[ht]
    \centering
    \includegraphics[width=0.85\linewidth]{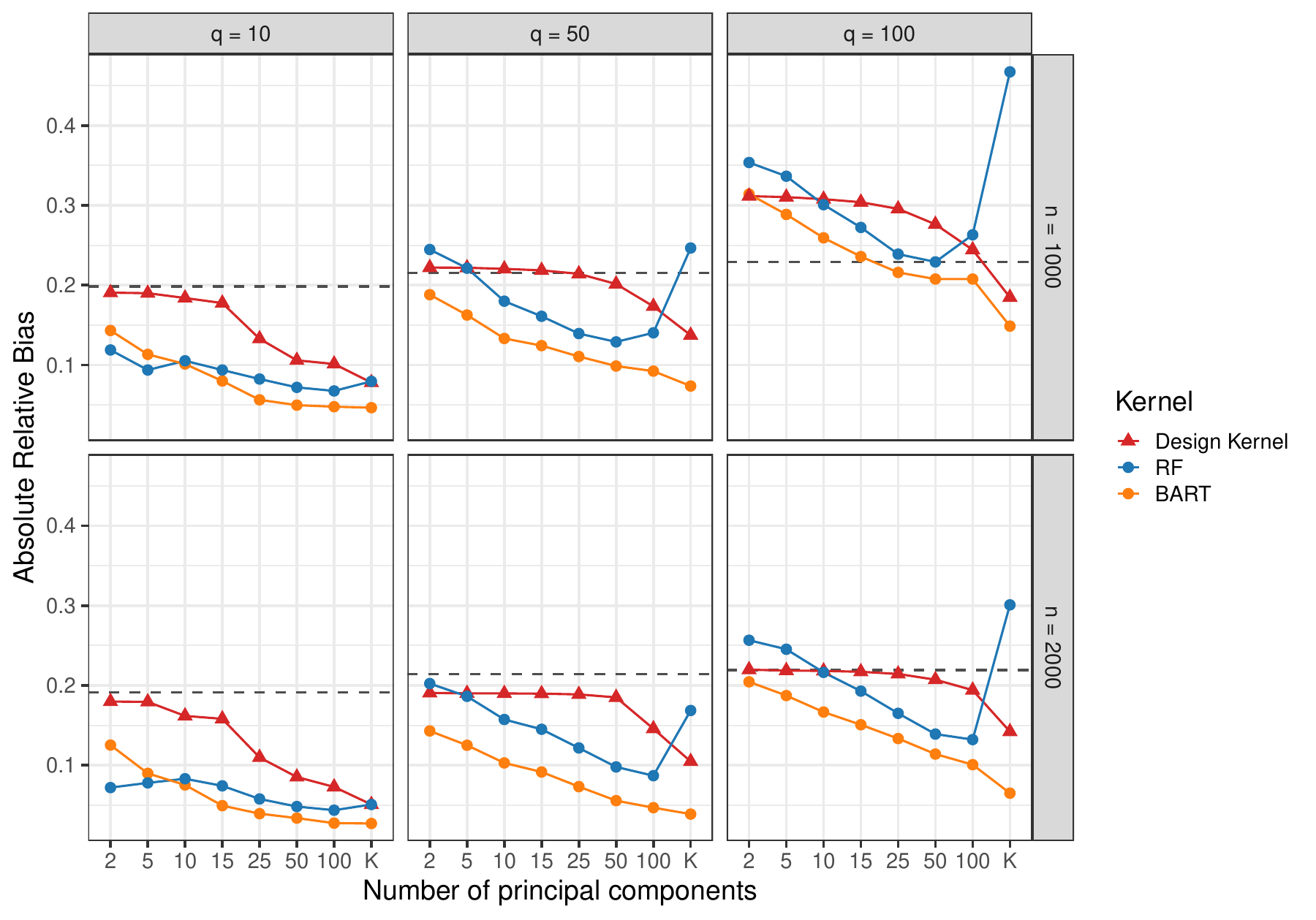}
    \caption{Relative bias for design-based and forest kernels, varying the covariate dimension $q$ and the analysis sample size $n$. Columns correspond to the number of covariates ($q = 10, 50, 100$) and rows to the analysis sample size ($n = 1000, 2000$). All estimates balance on both raw covariates and kernel features, varying the number of principal components on the horizontal axis, where $K$ denotes the full kernel. The black dashed line indicates performance from using raw covariates alone.}
    \label{fig:sim1-revision-bias}
\end{figure}

Figures \ref{fig:sim1-revision-rmse} and \ref{fig:sim1-revision-bias} plot relative RMSE and bias, respectively, against the number of balanced principal components, with columns indexing $q$ and rows indexing $n$. Both figures adjust for both raw covariates and kernel features. Like the simulation in the main text, the forest kernels maintain strong performance over the design-based kernel for moderate spectral truncations. As $q$ grows, the design-based kernel flattens and converges toward simple covariate adjustment (black dashed line) while the forest kernels continue to achieve lower bias and RMSE. Because the design-based kernel weights every covariate equally in its similarity metric, adding extra noise covariates dilutes the relevant signal, mirroring the outcome-agnostic behavior observed in Section \ref{sec:soldiering}. We also observe that adjusting for the entire random forest kernel results in poor performance, likely reflecting the use of deep trees in the random forest which produces a kernel that encodes fine, sample-specific partition structure; balancing on its full rank over-constrains the weights. The BART kernel, a sum of many shallow, weak trees, yields a smoother and denser kernel that remains stable at the full rank.

While absolute performance begins to degrade in high dimensions at the smaller sample size, increasing it to $n = 2000$ recovers the low-dimensional behavior, with the $q = 100$ panel closely resembling $q = 10$. The high-dimensional difficulty is therefore driven by the sample size relative to $q$ rather than a failure of the method: with enough data to estimate the kernel, forest kernel balancing continues to adjust for confounding effectively. As a practical matter, balancing on a moderate number of principal components, as recommended in the main text, avoids the full-rank regime where the random forest kernel is least stable.

\clearpage

\begin{revision}
\subsection{Weight estimation with CBPS} \label{sec:cbps}

In this section, we conduct an ablation study to assess whether the improvements from outcome-guided features depend on the choice of weighting procedure. We use the covariate balancing propensity score (CBPS) of \citet{imai2014covariate} as an alternative in the weighting step: given any set of input features, CBPS estimates a logistic propensity score model predicting treatment status given the features, subject to balance constraints. We repeat the simulation from Section \ref{sec:varying-q-n}, replacing the balancing weights with just-identified CBPS for the ATT.
We construct the features exactly as in Section \ref{sec:varying-q-n}. Because just-identified CBPS estimates one parameter per balanced feature, without regularization, it is not possible to fit CBPS weights on the full kernel ($K$), so we balance up to 100 principal components. Figures \ref{fig:cbps-sweep-rmse} and \ref{fig:cbps-sweep-bias} report the relative RMSE and bias, respectively.

Broadly, we find that the conclusions from Section \ref{sec:varying-q-n} carry over with this alternative choice of weighting procedure.
Using the learned forest features with CBPS outperforms using the raw covariates with CBPS for every combination of $q$ and $n$ at moderate spectral truncations.
We do find that the choice of weighting procedure somewhat affects performance in higher dimensional settings.
With $n=1000$ data points and $q=100$ raw features, the bias and RMSE using CBPS on the learned forest features are lower than when using CBPS with the raw features if the number of principal components is small to moderate, but performance decays rapidly with a higher number of PCs.
This is somewhat in contrast to the results in Section \ref{sec:varying-q-n} when $n=1000$ and $q=100$, where balancing weights with kernel features at first perform worse than balancing the raw features but improve with more PCs (the exception being using the learned RF features without spectral truncation, which performs poorly in the high dimensional low sample size setting, as we have discussed previously).
As in the main text, balancing on a moderate number of principal components avoids this instability.

\begin{figure}[ht]
    \centering
    \includegraphics[width=0.85\linewidth]{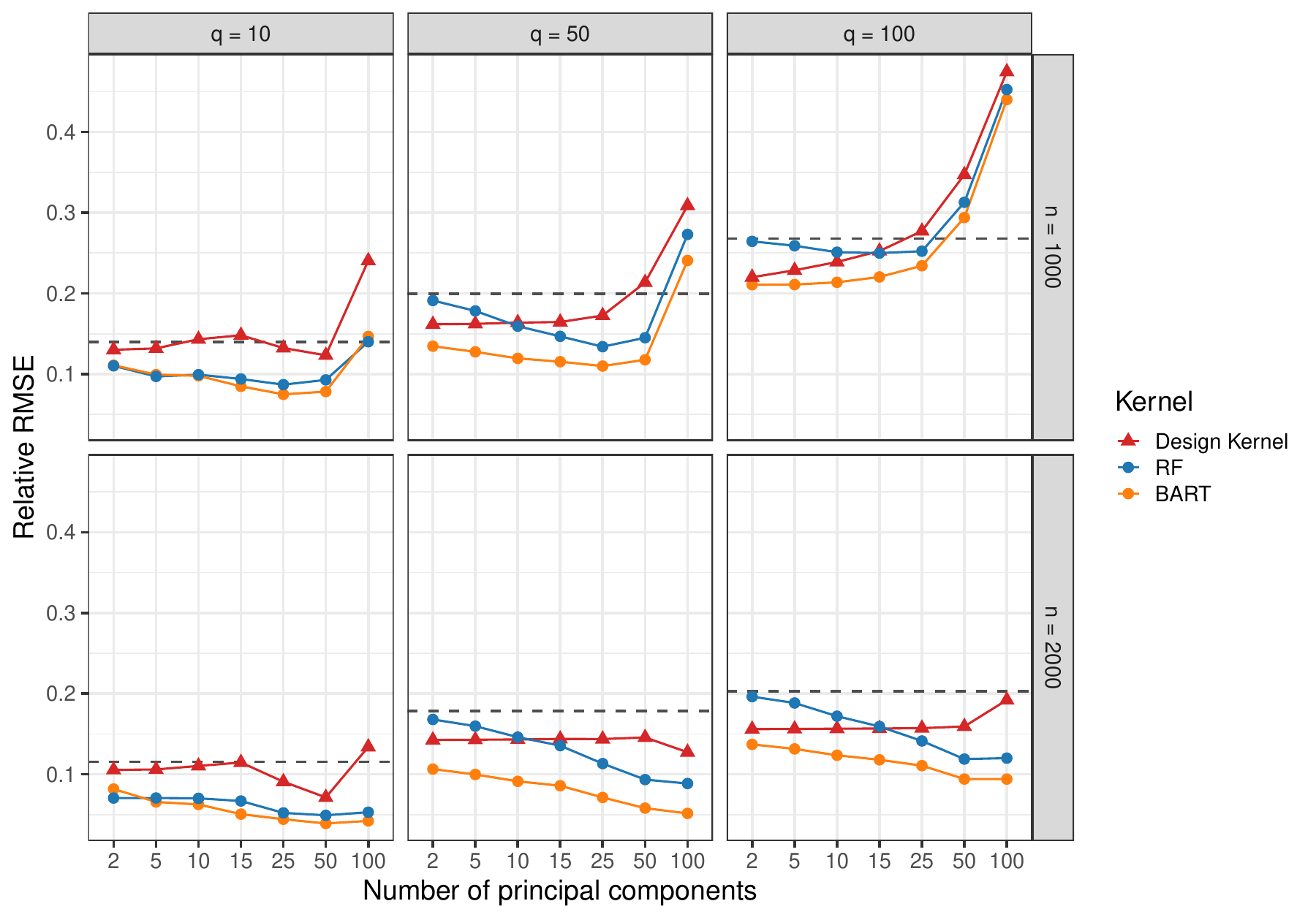}
    \caption{Relative RMSE for design-based and forest kernels with weights estimated by CBPS, varying the covariate dimension $q$ and the analysis sample size $n$. Columns correspond to the number of covariates ($q = 10, 50, 100$) and rows to the analysis sample size ($n = 1000, 2000$). All estimates balance on both raw covariates and kernel features, varying the number of principal components on the horizontal axis. The black dashed line indicates performance from CBPS applied to the raw covariates alone.}
    \label{fig:cbps-sweep-rmse}
\end{figure}

\begin{figure}[ht]
    \centering
    \includegraphics[width=0.85\linewidth]{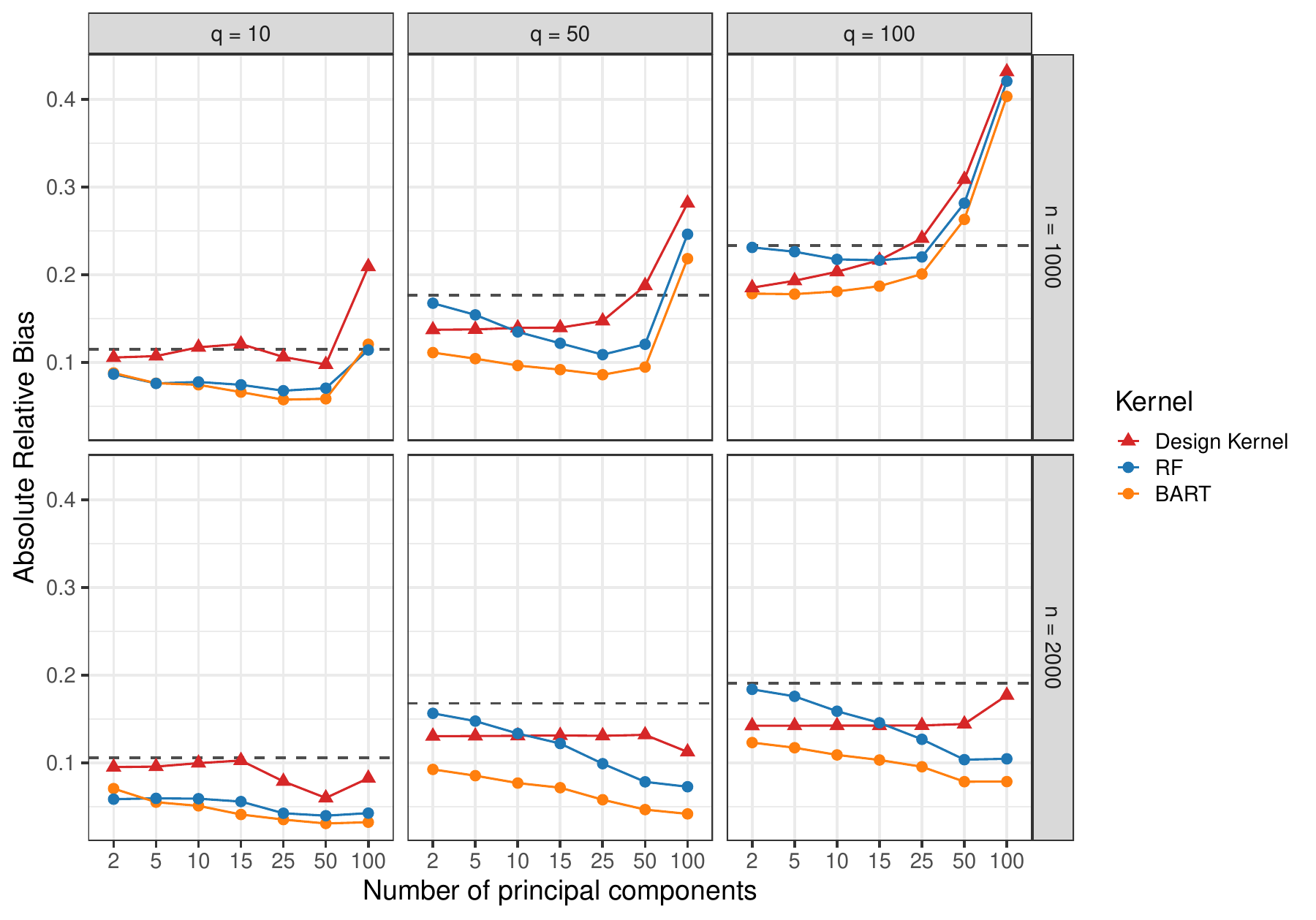}
    \caption{Relative bias for design-based and forest kernels with weights estimated by CBPS, varying the covariate dimension $q$ and the analysis sample size $n$. Columns correspond to the number of covariates ($q = 10, 50, 100$) and rows to the analysis sample size ($n = 1000, 2000$). All estimates balance on both raw covariates and kernel features, varying the number of principal components on the horizontal axis. The black dashed line indicates performance from CBPS applied to the raw covariates alone.}
    \label{fig:cbps-sweep-bias}
\end{figure}

\end{revision}

\clearpage

\subsection{Balance metrics}
\begin{figure}[ht]
    \centering
    \includegraphics[width=0.85\linewidth]{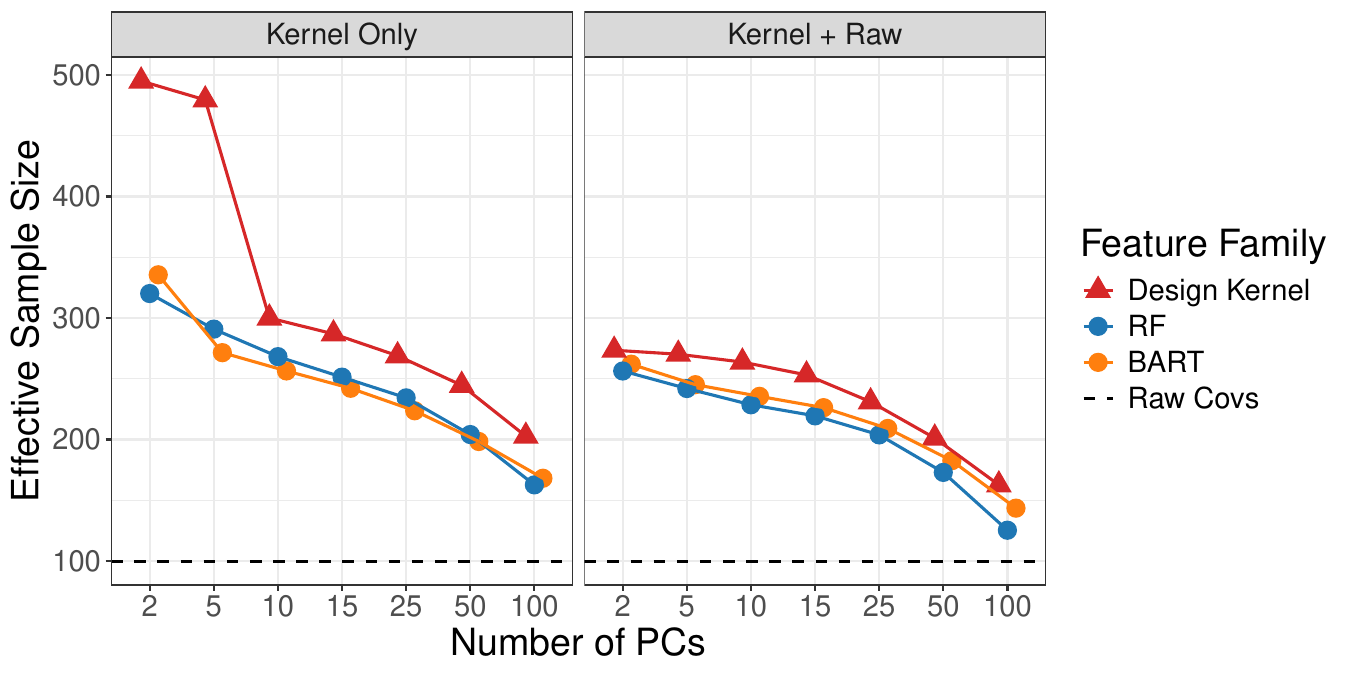}
    \caption{Effective sample size for design-based and forest kernels, varying the number of principal components to balance on. The left panel shows results using only kernel features (principal components); the right panel includes both raw features and kernel features. The black dashed line indicates ESS from using raw covariates alone.}
    \label{fig:sim1-ess}
\end{figure}

Figure \ref{fig:sim1-ess} displays the effective sample sizes (ESS) for the first simulation study. Discussion of this metric can be found in Section 4 of the main manuscript.

\subsection{Weight estimation with logistic regression}
\begin{figure}[ht]
    \centering
    \includegraphics[width=0.85\linewidth]{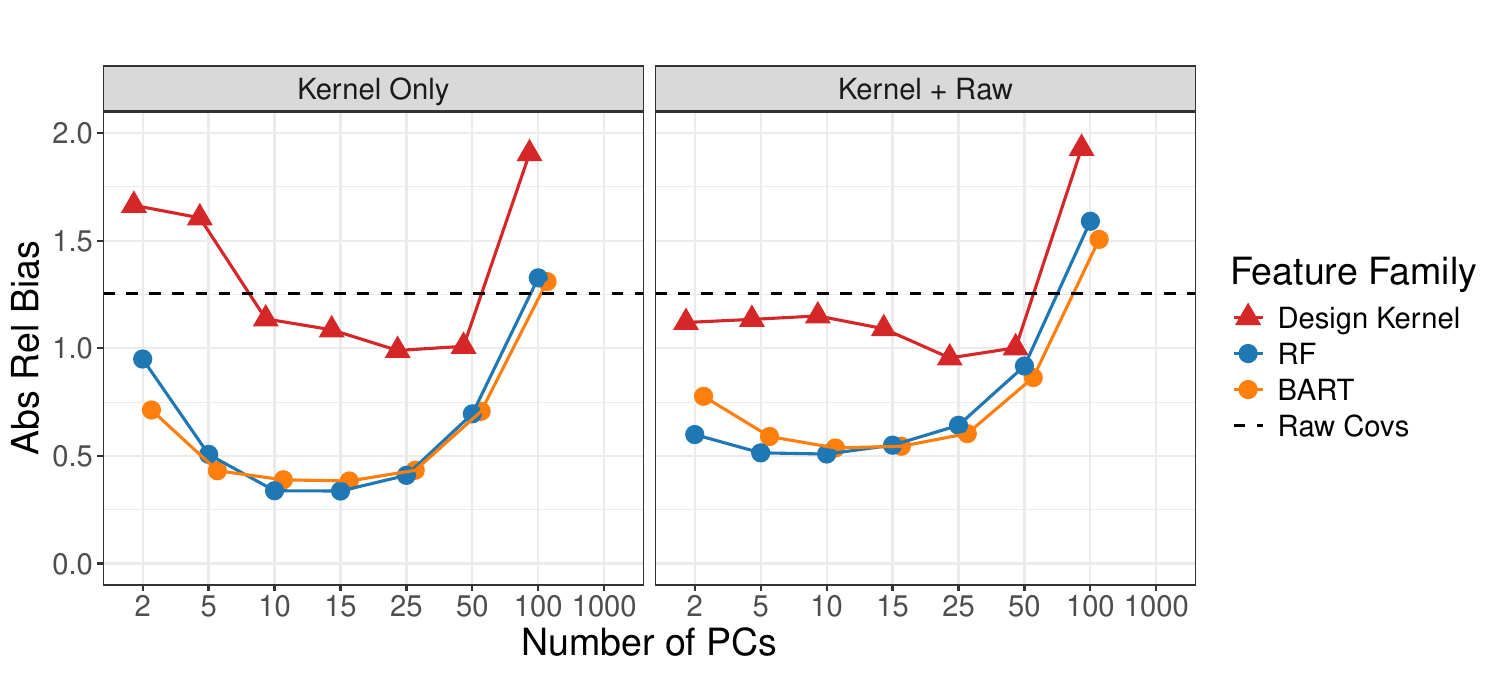}
    \caption{Results when logistic regression is used for weight estimation instead of balancing weights. We plot absolute relative bias for design-based and forest kernels, varying the number of principal components to balance on. The left panel shows results using only kernel features (principal components); the right panel includes both raw features and kernel features. The black dashed line indicates performance from using raw covariates alone. }
    \label{fig:sim1-relbias-ipw}
\end{figure}

\begin{figure}[ht]
    \centering
    \includegraphics[width=0.85\linewidth]{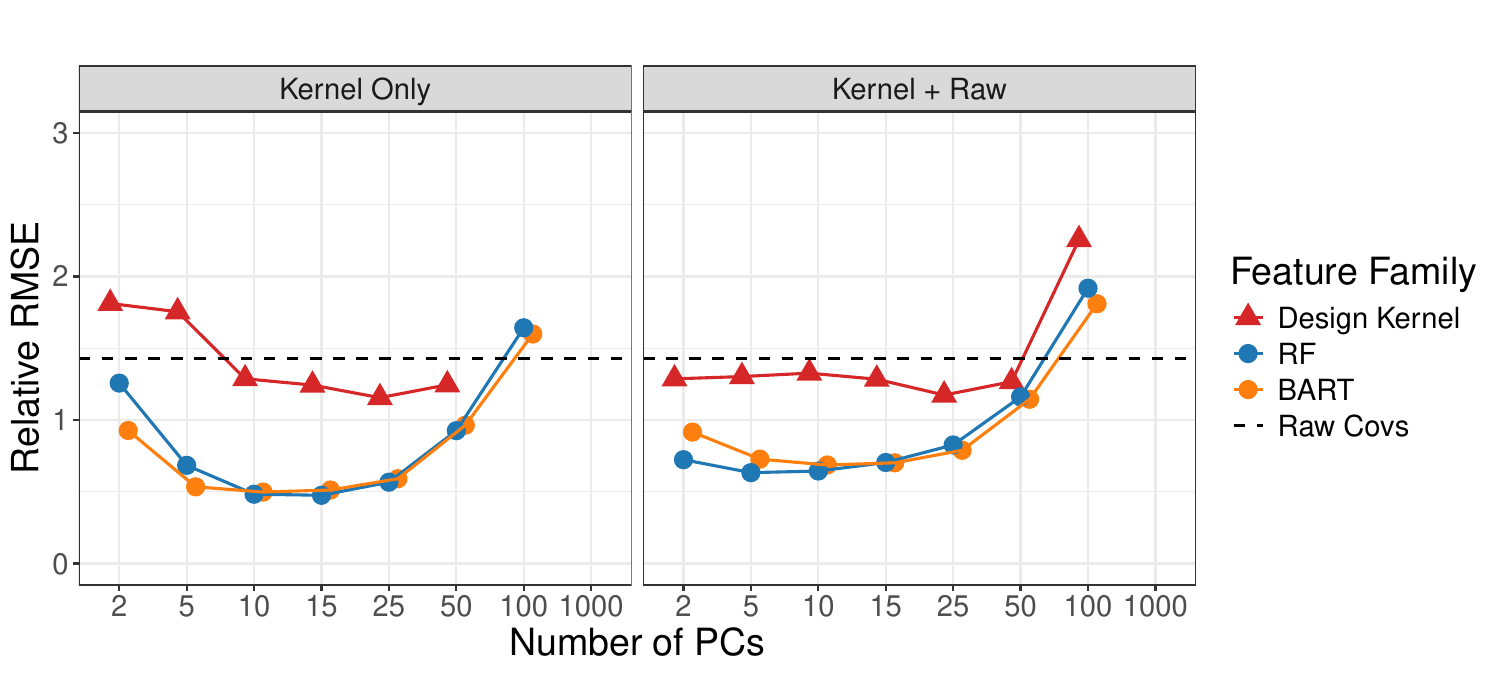}
    \caption{Results when logistic regression is used for weight estimation instead of balancing weights. We plot absolute relative RMSE for design-based and forest kernels, varying the number of principal components to balance on. The left panel shows results using only kernel features (principal components); the right panel includes both raw features and kernel features. The black dashed line indicates performance from using raw covariates alone. }
    \label{fig:sim1-relrmse-ipw}
\end{figure}

In Figures \ref{fig:sim1-relbias-ipw} and \ref{fig:sim1-relrmse-ipw}, we present results from the first simulation using logistic regression to estimate the propensity score instead of direct estimation with balancing weights. The substantive conclusions remain the same as those in the main manuscript, indicating that the method for weight estimation is not critical in the forest kernel balancing framework.

\clearpage

\section{Additional results from child soldiering study} \label{sec:extra-soldiering-results}

\begin{figure}[ht]
    \centering
    \includegraphics[width=0.8\linewidth]{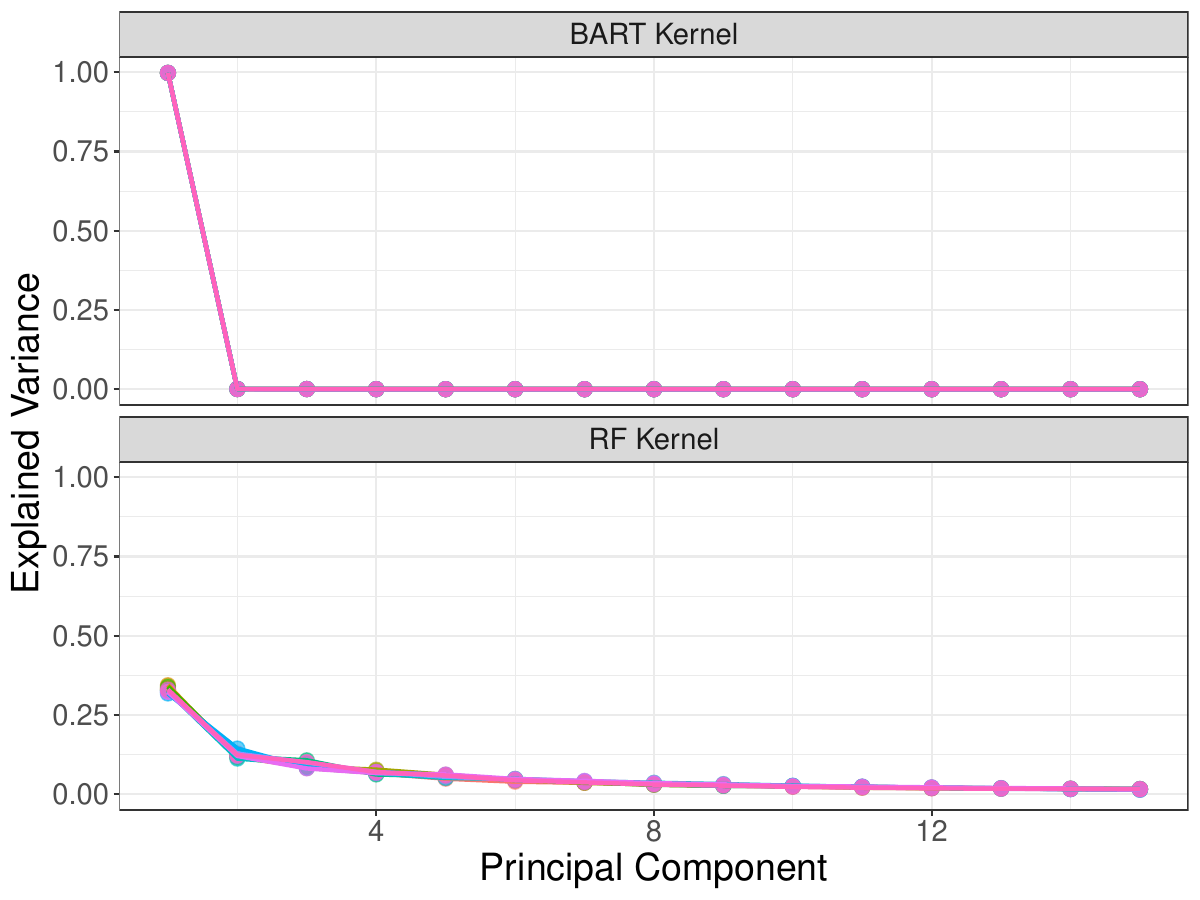}
    \caption{Scree plot of principal components for BART and RF kernels for child soldiering study for all ten cross-fit folds. We plot the principal component on the horizontal and the corresponding explained variance on the vertical axis.}
    \label{fig:scree-soldiering}
\end{figure}

\section{Additional results from within-study comparison} \label{sec:extra-wsc-results}

\begin{figure}[ht]
    \centering
    \includegraphics[width=0.95\linewidth]{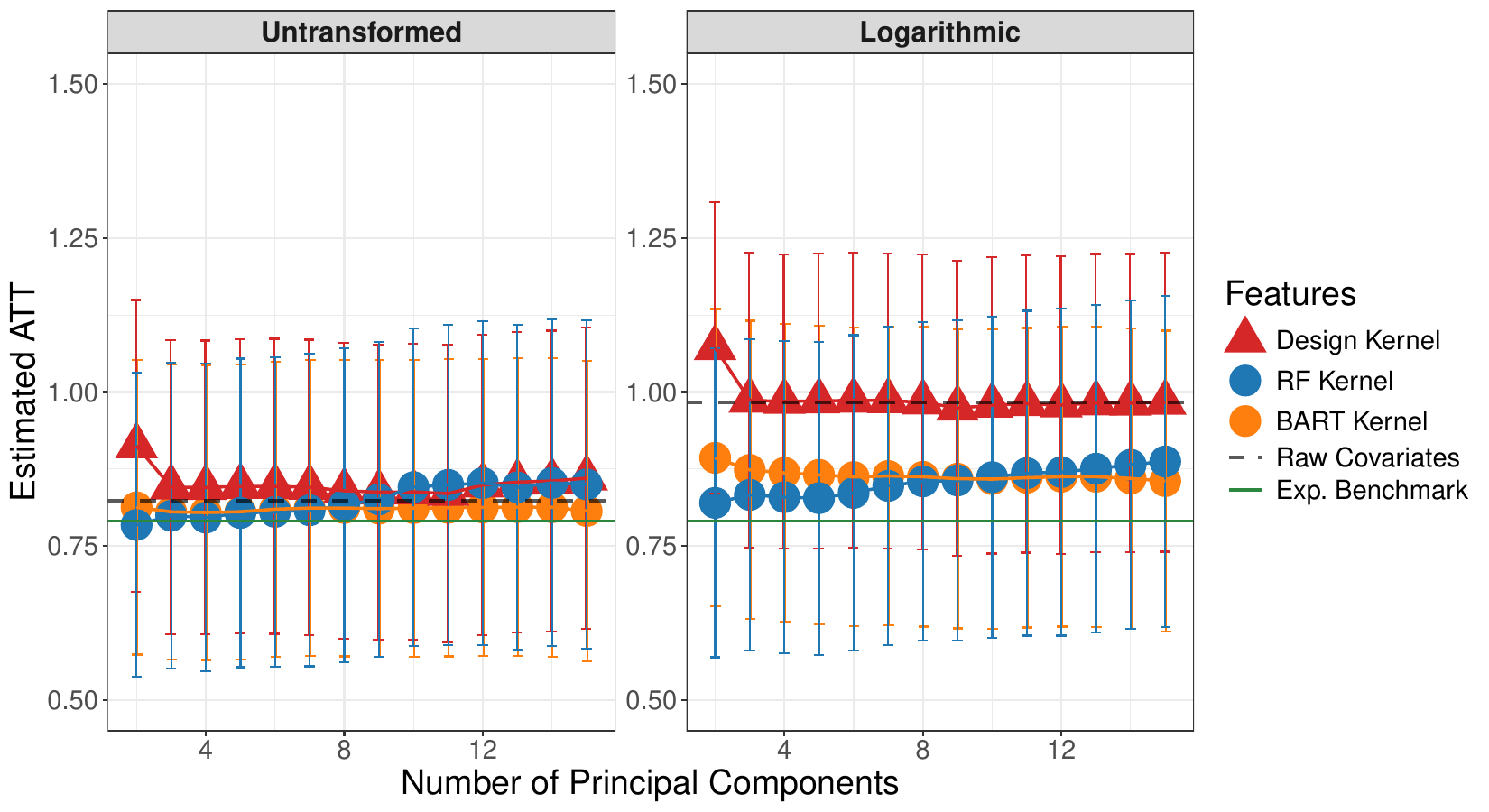}
    \caption{Full results for WSC empirical study with logarithmic ($\log(x+1)$) covariate transformation. We plot the estimated ATT over a range of principal components (PCs) for kernel balancing methods. All points adjust for raw covariates and the specified number of PCs. The black dashed line is the estimate from using raw covariates only. The green solid line is the experimental benchmark.}
    \label{fig:wsc-full-log}
\end{figure}

\begin{figure}[ht]
    \centering
    \includegraphics[width=0.95\linewidth]{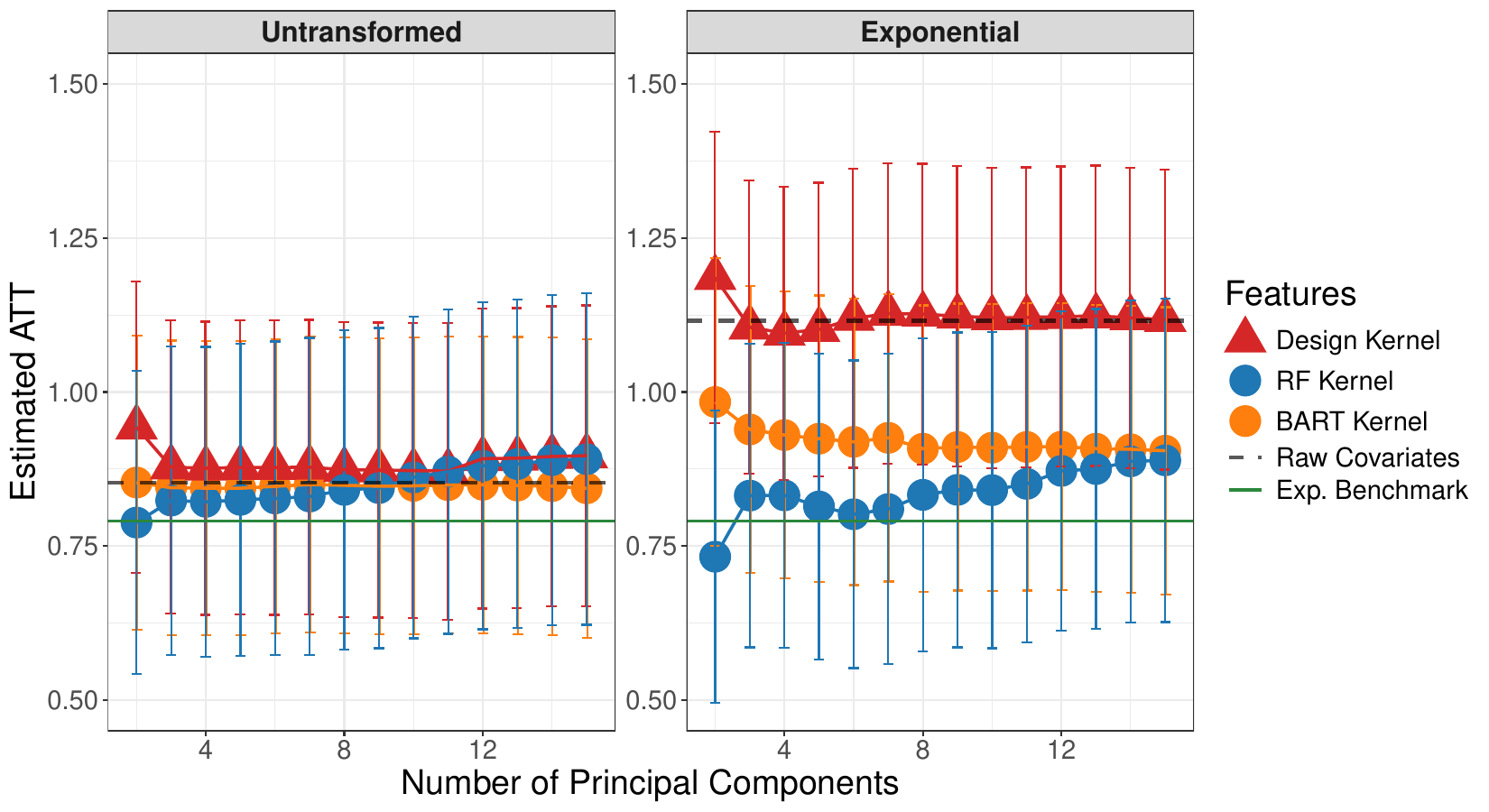}
    \caption{Full results for WSC empirical study with exponential ($\exp(x)$) covariate transformation. We plot the estimated ATT over a range of principal components (PCs) for kernel balancing methods. All points adjust for raw covariates and the specified number of PCs. The black dashed line is the estimate from using raw covariates only. The green solid line is the experimental benchmark.}
    \label{fig:wsc-full-exp}
\end{figure}

\begin{figure}[ht]
    \centering
    \includegraphics[width=0.8\linewidth]{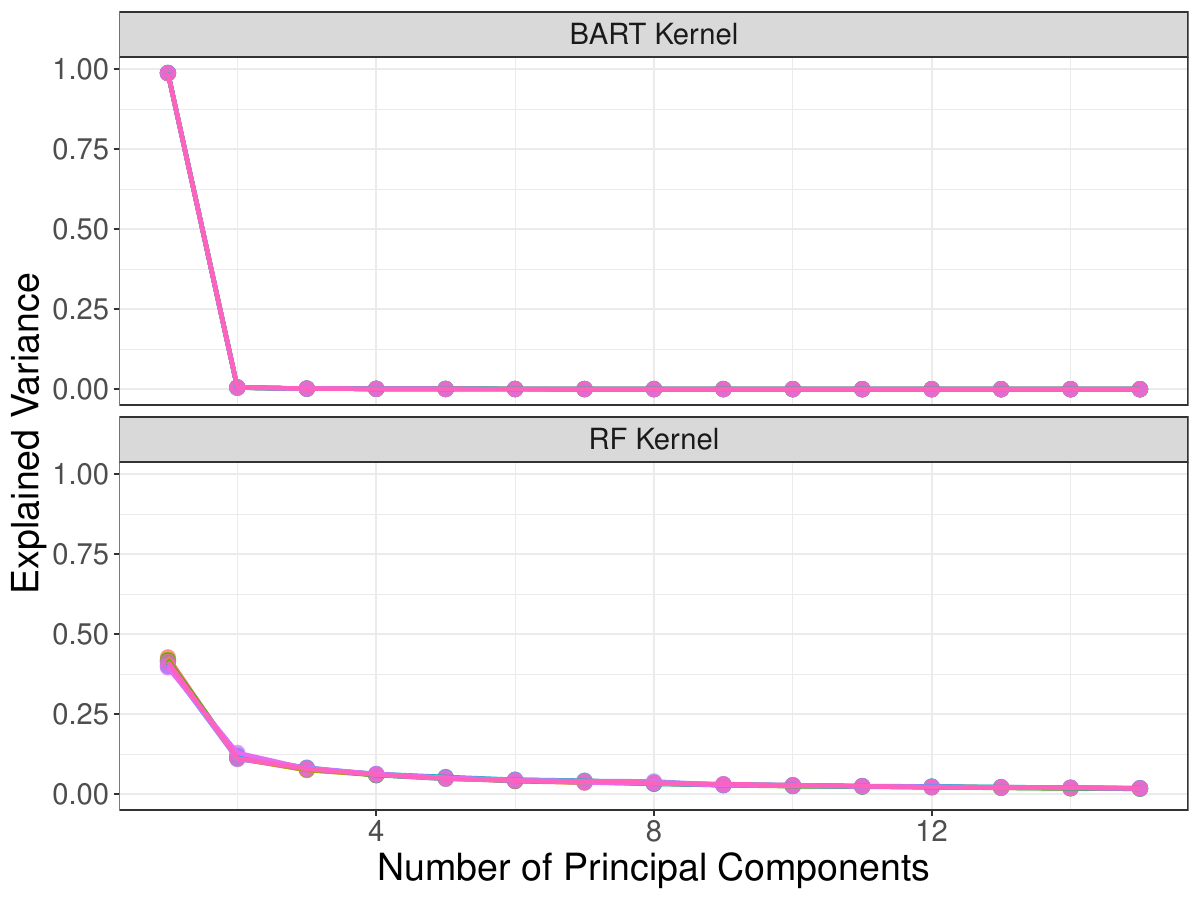}
    \caption{Scree plot of principal components for BART and RF kernels for within-study comparison for all ten cross-fit folds. We plot the principal component on the horizontal and the corresponding explained variance on the vertical axis.}
    \label{fig:scree-wsc}
\end{figure}

\clearpage

\section{Additional simulation study} \label{sec:extra-simulations}

\subsection{Simulation setup}
The second simulation is adapted from \citet{kim2024scalable}.  This DGP also represents a highly nonlinear scenario in both the propensity score and the outcome model. In addition, we also control the overlap between treated and control groups (defined below).

In this simulation, we construct the following observed covariates: 
\begin{align*}
    &\begin{bmatrix}
X_1 \\
X_2 \\
X_3
\end{bmatrix}
\sim N
\left(
\begin{bmatrix}
0 \\
0 \\
0
\end{bmatrix},
\begin{bmatrix}
2 & 1 & -1 \\
1 & 1 & -0.5 \\
-1 & -0.5 & 1
\end{bmatrix}
\right), \quad X_4 \sim \text{Unif}[-3,3], \quad X_5 \sim \chi_1^2, \quad X_6 \sim \text{Bern}[0.5].
\end{align*}

The treatment $Z$ and outcome $Y$ are generated as
\begin{align*}
    Z_i &= \mathbbm{1} \left\lbrace X_{i1}^2 + 2X_{i2}^2 - 2X_{i3}^2 - (X_{i4} + 1)^3 - 0.5 \log(X_{i5} + 10) + X_{i6} - 1.5 + \varepsilon_i > 0 \right\rbrace, \quad \varepsilon_i \sim \mathcal{N}(0, \sigma_{\varepsilon}^2),\\
    Y_i &= (X_{i1} + X_{i2} + X_{i5})^2 + \eta_i, \quad \eta_i \sim \mathcal{N}(0,1).
\end{align*}
Following \citet{kim2024scalable}, we consider $\sigma_{\varepsilon}^2 = 30$ and $\sigma_{\varepsilon}^2 = 100$, corresponding to the \textit{low} and \textit{high overlap} settings, respectively. Observe that the ATT is equal to 0 by construction.

\begin{figure}[ht]
    \centering
    \includegraphics[width=1\linewidth]{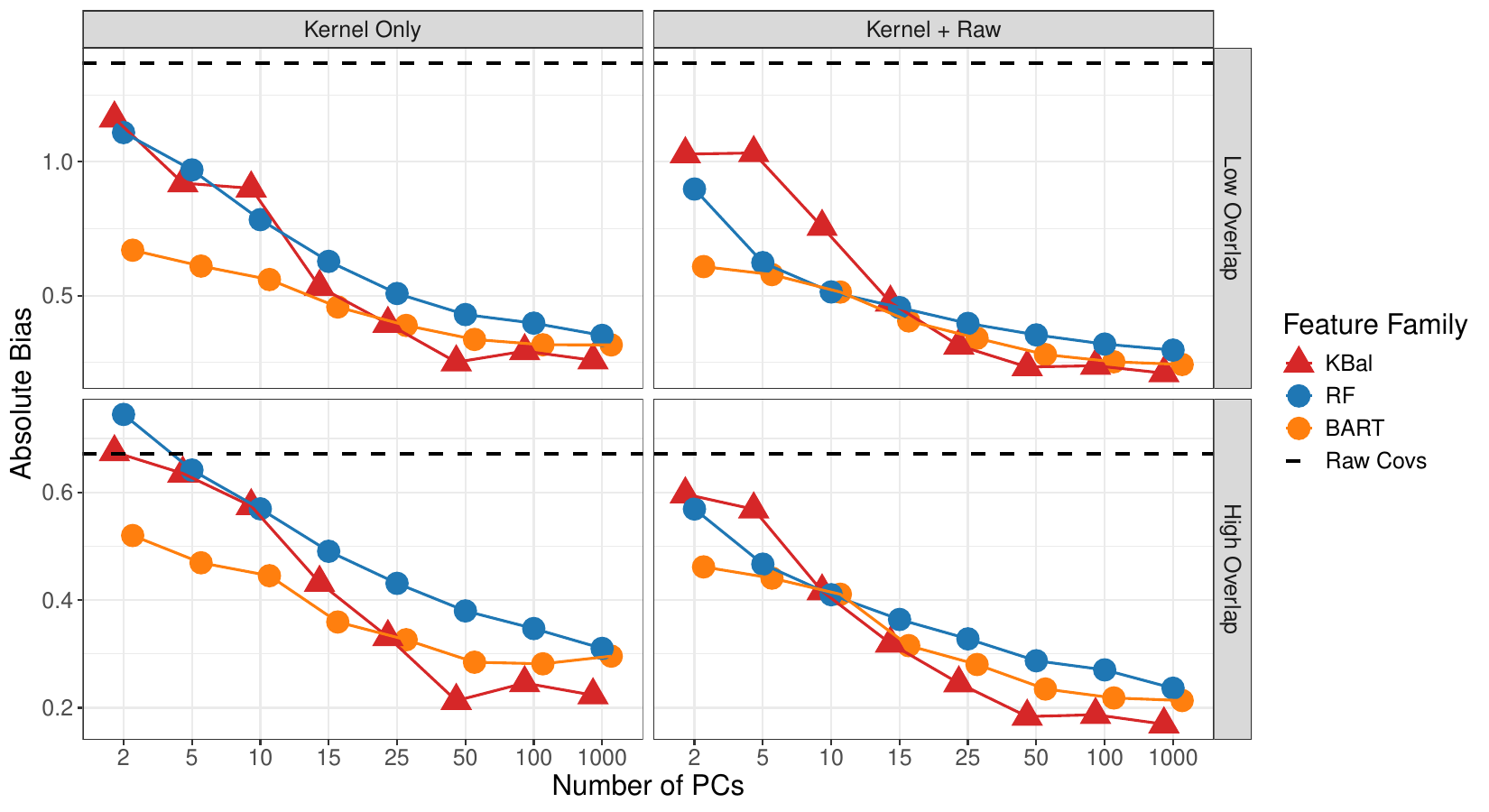}
    \caption{Bias plot for design-based and forest kernels, varying the number of principal components to balance on. The left panel shows results using only kernel features (principal components); the right panel includes both raw features and kernel features. The black dashed line indicates performance from using raw covariates alone.}
    \label{fig:sim2-bias}
\end{figure}

\begin{figure}[ht]
    \centering
    \includegraphics[width=1\linewidth]{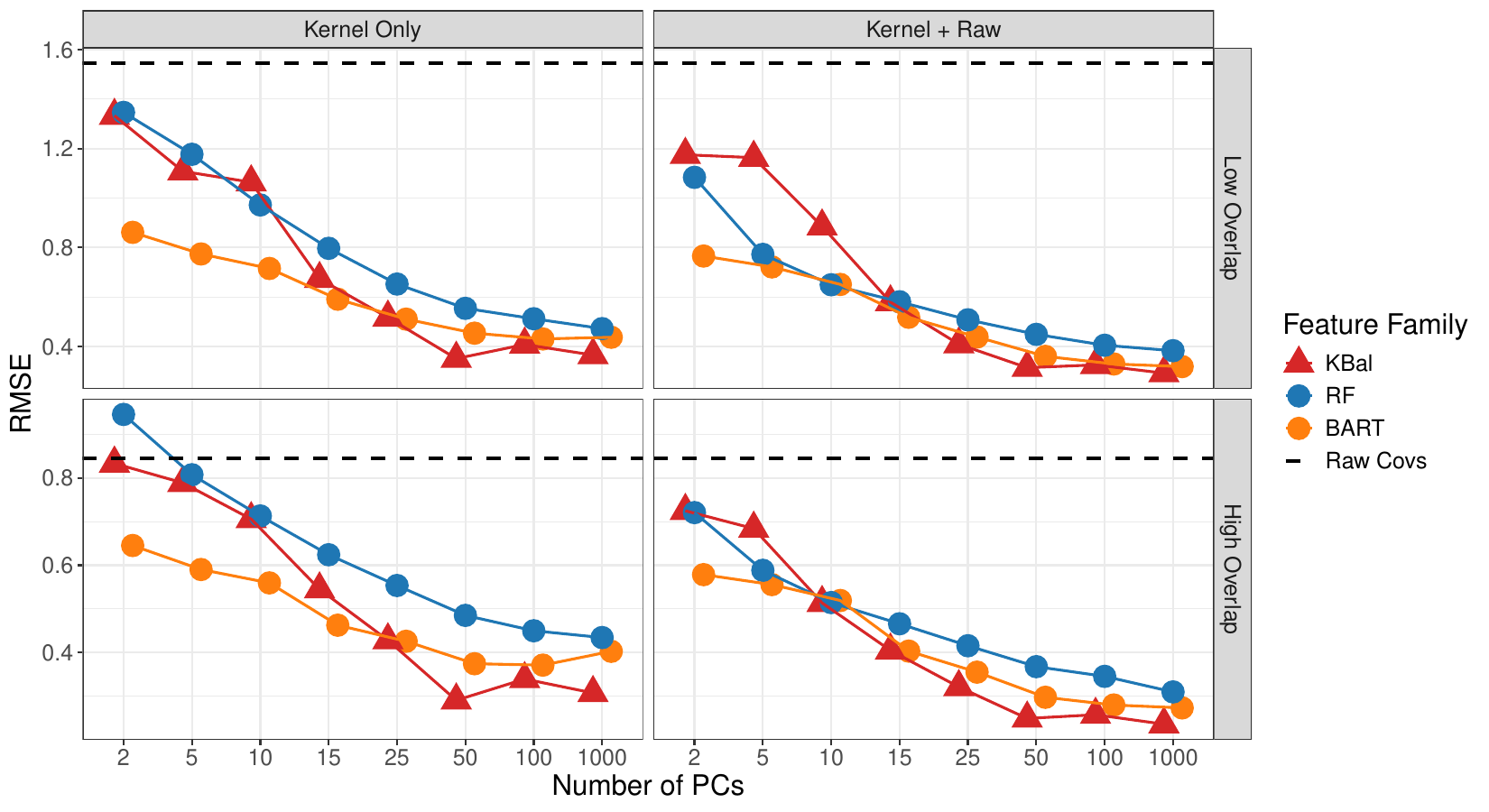}
    \caption{RMSE plot for design-based and forest kernels, varying the number of principal components to balance on. The left panel shows results using only kernel features (principal components); the right panel includes both raw features and kernel features. The black dashed line indicates performance from using raw covariates alone.}
    \label{fig:sim2-rmse}
\end{figure}

Following the setup in the main manuscript, we compare the performance of forest kernel balancing weights against a design-based Gaussian kernel and balancing on raw covariates only. The results are included in Figures \ref{fig:sim2-bias} (bias) and \ref{fig:sim2-rmse} (RMSE).

Figures \ref{fig:sim2-bias-ipw} and \ref{fig:sim2-rmse-ipw} below display results for this simulation study when logistic regression is used for weight estimation instead of balancing weights.

\begin{figure}[ht]
    \centering
    \includegraphics[width=1\linewidth]{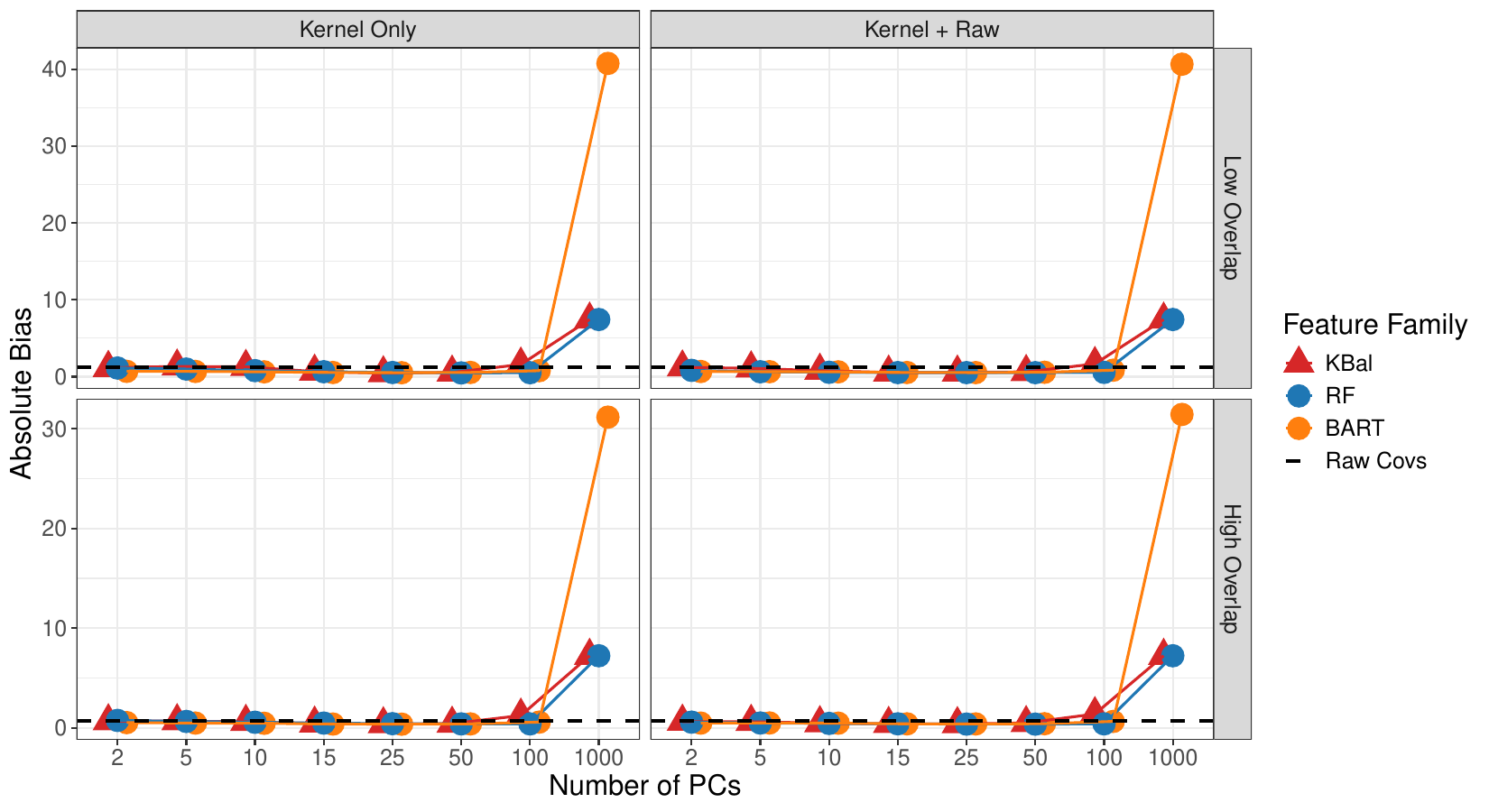}
    \caption{Results when logistic regression is used for weight estimation instead of balancing weights. We plot absolute bias for design-based and forest kernels, varying the number of principal components to balance on. The left panel shows results using only kernel features (principal components); the right panel includes both raw features and kernel features. The black dashed line indicates performance from using raw covariates alone.}
    \label{fig:sim2-bias-ipw}
\end{figure}

\begin{figure}[ht]
    \centering
    \includegraphics[width=1\linewidth]{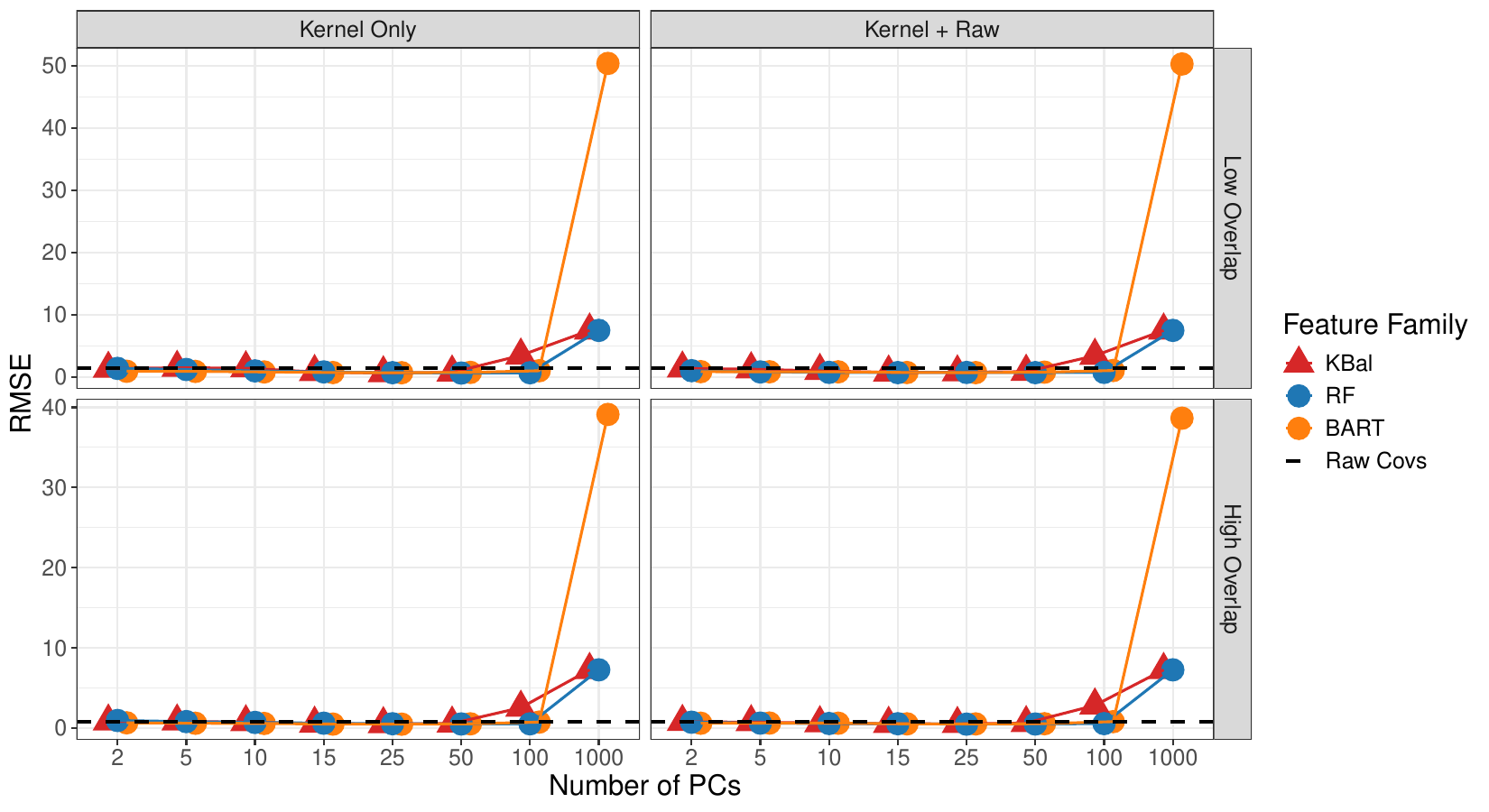}
    \caption{Results when logistic regression is used for weight estimation instead of balancing weights. We plot RMSE for design-based and forest kernels, varying the number of principal components to balance on. The left panel shows results using only kernel features (principal components); the right panel includes both raw features and kernel features. The black dashed line indicates performance from using raw covariates alone.}
    \label{fig:sim2-rmse-ipw}
\end{figure}


\end{document}